\documentclass{article}
\usepackage[margin=2cm]{geometry}
\usepackage{amsmath, amssymb, amsfonts}
\usepackage{graphicx, subcaption}
\usepackage{multirow}
\usepackage{tikz}
\usetikzlibrary{positioning,arrows,chains,matrix,scopes,fit,decorations.markings,decorations}
\tikzstyle{block}=[draw opacity=0.7,line width=1.4cm]

\newcommand\blfootnote[1]{%
	\begingroup
	\renewcommand\thefootnote{}\footnote{#1}%
	\addtocounter{footnote}{-1}%
	\endgroup
}

\numberwithin{equation}{section}
\newcommand{\prf}{\underline{Proof:}\ }
\newcommand{\finprf}{\null
  \hfill {\rule{5pt}{5pt}}}
\newcounter{NN}
\newtheorem{proposition}[NN]{Proposition}
\newtheorem{theorem}[NN]{Theorem}
\newtheorem{corollary}[NN]{Corollary}

\newtheorem{lemma}[NN]{Lemma}
\newtheorem{definition}[NN]{Definition}

\newcommand{\bma}{\begin{pmatrix}} \newcommand{\ema}{\end{pmatrix}}
\newcommand{\ie}{{i.e.}\ }

\def\L{{\mathcal L}}

\def\E{{\mathcal E}}

\def\id{{\text{id}}}

\newcommand{\ZZ}{{\mathbb Z}}

\newcommand{\sla}{{\sigma(\lambda)}}

\title{Integrable boundary conditions for quad equations,  \\ open boundary reductions and integrable mappings}
\date{\empty}
\author{Vincent Caudrelier\blfootnote{email addresses: v.caudrelier@leeds.ac.uk; ch.zhang.maths@gmail.com; p.vanderkamp@latrobe.edu.au}$^{1}$, Peter H.~van der Kamp$^{2}$, Cheng Zhang$^{3}$ \\[2mm]
  \sc $^{1}$\small School of Mathematics\\
   \sc \small University of Leeds, LS2 9JT Leeds, UK \\[1mm]
  \sc $^{2}$\small Department of Mathematics and Statistics \\ \sc \small La Trobe University,     Victoria 3086, Australia\\[1mm]
  \sc $^{3}$\small Department of Mathematics \\
  \sc \small Shanghai University, Shanghai 200444, China
 }

\begin{document}

\maketitle

\begin{abstract}
In the context of integrable partial difference equations on quad-graphs, we introduce the notion of {\em  open boundary reductions} as a new means to construct discrete integrable mappings and their invariants. This represents an alternative to the well-known periodic reductions. The construction deals with well-posed initial value problems for quad equations on quad-graphs restricted to a strip. It relies on the so-called {\em double-row monodromy matrix} and gives rise to integrable mappings. To obtain the double-row monodromy matrix, we use the notion of {\em boundary matrix} and {\em  discrete boundary zero curvature condition}, themselves related to the boundary consistency condition, which complements the well-known $3$D consistency condition for integrable quad equations and gives an integrability criterion for boundary equations. This relation is made precise in this paper.
Our focus is on quad equations from the Adler-Bobenko-Suris (ABS) classification and their discrete integrable boundary equations. Taking as our prime example a regular $\ZZ^2$-lattice with two parallel boundaries, we provide an explicit construction of the maps obtained by open boundary reductions, the boundary matrices, as well as the invariants extracted from the double-row monodromy matrix. Examples of integrable maps are considered for the H1 and Q1($\delta =0$) equations and an interesting example of a non-QRT map of the plane is presented. Examples of well-posed quad-graph systems on a strip beyond the $\ZZ^2$-lattice are also given.
\vspace{1mm}


\end{abstract}

\section{Introduction}
Some of the early motivations for integrable ordinary and partial difference equations dealt with the space and/or time discretization of integrable ordinary/partial differential equations with the view of exact numerical schemes. This area is now a well-established field of its own with ramifications well beyond the original motivations, see for instance the monographs \cite{BSbook, Dui, HJN}. %A theory of discrete integrability has emerged.
The concept of commuting flows in the continuous world finds an analogue in the discrete world known as multidimensional consistency (or consistency around the cube) \cite{BS,nijhoff2002lax, NW1} for a certain class of partial difference equations formulated on quad-graphs. There is a deep connection between this notion and that of the set-theoretical Yang-Baxter equation \cite{Drin}. Many other important features of integrability, such as Lax pairs and B\"acklund transformations, are shared between the continuous and discrete worlds.

More recently, the idea of a boundary consistency condition
%to associate integrable boundary equations to integrable quad equations
was introduced in \cite{CCZ}, and emerged from the introduction of the set-theoretical reflection equation \cite{CCZ1, CZ}. The latter is a companion of the set-theoretical Yang-Baxter equation, and its solutions are called reflections maps, in analogy with Yang-Baxter maps \cite{veselov2003yang}. Similarly, the boundary consistency condition is a companion to the multidimensional consistency condition for quad equations, and is used to define associated discrete integrable boundary equations.

Integrable maps are the discrete analog of integrable Hamiltonian flows, and are thus of central importance, cf. \cite{Ves}. They have been well studied and many examples have been constructed over the years, e.g. the McMillan map \cite{McMillan} and the QRT maps \cite{QRT1, QRT2}. A systematic scheme to obtain higher-dimensional discrete maps with (potentially) enough invariants to ensure (discrete) Liouville integrability \cite{Ves} is the periodic reduction method, also known as the staircase method \cite{PNC,KQ}. It relies on the possibility to define a well-posed initial value problem for partial difference equations on quad-graphs \cite{AV,VdK2} (typically in the form of a staircase for the $\ZZ^2$-lattice case \cite{VdK1}), and to impose periodicity conditions on this initial value problem. The construction in \cite{PNC} employs the Lax pair of the partial difference equation at hand. Invariant functions are obtained from the associated monodromy matrix, which is a product of Lax-matrices along the staircase. The involutivity of the invariants and complete integrability were proved later in \cite{CNP}.

\begin{figure}[h!]
  \centering
  \begin{tikzpicture}[scale=2]
    \tikzstyle{nod1}= [circle, inner sep=0pt, minimum size=3pt, draw]
    \tikzstyle{nod}= [circle, inner sep=0pt, fill=black, minimum size=3pt, draw]
    \tikzstyle{vertex}=[circle,minimum size=20pt,inner sep=0pt]
    \tikzstyle{selected vertex} = [vertex, fill=red!24]
    \tikzstyle{selected edge} = [draw,line width=1.5pt,-]
    \tikzstyle{edge} = [draw, thin,-,black] %inner lines
    \tikzstyle{ddedge} = [draw, densely dotted,-,black] %dotted lines
    \tikzstyle{dedge} = [draw, dashed,-,black] %dashed lines
    \tikzstyle{eedge} = [draw, thick,-,black] %outer lines

    \pgfmathsetmacro \l {.02}% inner lenght
    % 1st floor v, v1, v4, v14
    \node[nod1] (v1) at (4.5*\l,182.9*\l) {};

    \node[nod1] (v2) at (30*\l,183*\l) {};
    \node[nod1] (v3) at (55.5*\l,192.4*\l) {};
    \node[nod1] (v4) at (62*\l,216.5*\l) {};
    \node[nod1] (u1) at (25*\l,168*\l) {};
    \node[nod1] (u2) at (51*\l,171*\l) {};
    \node[nod1] (u3) at (76*\l,177*\l) {};
    \node[nod1] (u4) at (82.4*\l,201*\l) {};
    \node[nod1] (w1) at (28*\l,142*\l) {};
    \node[nod1] (w2) at (55*\l,145*\l) {};
    \node[nod1] (w3) at (79*\l,152*\l) {};
    \node[nod1] (w4) at (104*\l,149*\l) {};
    \node[nod1] (x1) at (44*\l,122*\l) {};
    \node[nod1] (x2) at (69*\l,129*\l) {};
    \node[nod1] (x3) at (80.5*\l,105*\l) {};
    \node[nod1] (x4) at (85.5*\l,127*\l) {};
    \node[nod1] (x5) at (111*\l,125*\l) {};
    \node[nod1] (y4) at (107*\l,198*\l) {};
    \node[nod1] (y7) at (105*\l,174*\l) {};

    \draw[thick] (v1) -- (v2) -- (v3)    --   (v4) ;
    \draw[thick] (u1) -- (u2) -- (u3)    --   (u4) ;
    \draw[thick] (w1) -- (w2) -- (w3)    --   (w4) ;
    \draw[thick] (x1) -- (x2) -- (x3)    --   (x4) -- (x5) -- (x3) ;
    \draw[thick] (v1) -- (u1) -- (w1)    --   (v1) ;
    \draw[thick] (v2) -- (u2) -- (w2);
    \draw[thick] (v3) -- (u3) -- (w3) --(x2);
    \draw[thick] (v4) -- (u4) -- (y4);
    \draw[thick] (w3) --(x4);
    \draw[thick] (w1) -- (x1) -- (w2) ;
    \draw[thick] (x1) -- (x3);
    \draw[thick] (u3) -- (y7) -- (y4);
    \draw[thick] (y7) -- (w4) -- (x5);
    \draw[line width=1.5pt] (v1) -- (w1) -- (x1) -- (x3) -- (x5);
\end{tikzpicture}
  \caption{A quad-graph with a boundary consisting of triangular faces} \label{fig:bccy1}
\end{figure}
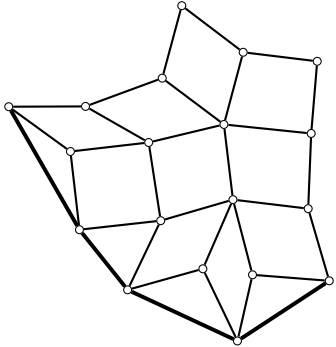

In this paper, we present a new scheme to produce integrable maps as well as a generating function for their invariants. It is based on the notion of integrable boundary conditions for partial difference equations on quad-graphs with a boundary. The boundary on a quad-graph naturally consists of triangular faces on which one defines a boundary equation, as is shown in Figure~\ref{fig:bccy1}. The boundary equations are said to be integrable if the associated boundary consistency condition is satisfied \cite{CCZ}.

A key tool in our construction of invariants is the {\it double-row monodromy matrix}, originally introduced by Sklyanin \cite{sklyanin1987boundary,sklyanin1988boundary}, whose properties depend on the representation of integrable boundary equations via the {\em discrete boundary zero curvature equation} involving a {\em boundary matrix}.

We call the reductions obtained by our method {\it open boundary reductions}, in analogy, for instance,  with the open boundary problems for quantum spin chains \cite{sklyanin1988boundary}. They are constructed from a well-posed initial value problem on a quad-graph with boundary where integrable boundary equations are imposed.

The paper is organized as follows: in Section $2$, we review and clarify the relevant ingredients involved in discrete integrable boundary conditions for equations on a quad-graph with a boundary. The key notion, introduced in \cite{CCZ}, is the boundary consistency condition for the boundary equation. We show how this condition is related to a second (dual) boundary consistency condition and subsequently to a zero curvature condition for the boundary equation. An important tool in doing so is the idea of folding explained in Section \ref{folding}. Special emphasis is given to the ABS list \cite{adler2003classification} and its associated boundary equations. In Section \ref{open_red}, we construct (non-)autonomous mappings as open boundary reductions of partial difference equations, from well-posed initial value problems on the $\ZZ^2$-lattice with two parallel boundaries. The boundary zero curvature representation of the integrable boundary conditions ensures (under certain conditions) the (anti-)isospectral property of the double-row monodromy matrix. This enables us to construct a generating function for (2-)integrals of the mappings. More generally, our construction also provides us with $k$-integrals for $(k+1)$-dimensional mappings, cf. \cite{HBQC}. In Section \ref{examples}, we illustrate the results of Section \ref{open_red} by explicitly computing some low-dimensional mappings and their invariants. 
Interestingly, the 3-dimensional non-autonomous open reduction of the Q1($\delta =0$) equation gives rise to the following 2-parameter family of {\it non-QRT} planar maps:
\begin{equation}
(x,y)\mapsto \frac { \left( x+y \right)  \left(
\alpha x + \beta(\alpha x + y+1)y
\right) ^{2}}{ \beta\, \left(
x+(x\beta+y+1)y
\right) \left(
(\alpha^{2}+\beta)x{y}^{2}
+\alpha(\beta{x}^{2}+{y}^{2})y
+\alpha(x+y)^2
\right) }\left({\frac {x \left(
\alpha x+ (\alpha\beta x + \beta y +\alpha)y
\right) }{\alpha \left(
x+(\alpha x+y+1)y
\right) }},y\right).
\end{equation}
It leaves invariant the pencil of cubic curves of genus 0:
\begin{equation} \label{pencil}
 y^2 (1 + x + y) + \alpha x (\frac{x}{\beta} + x y +  y^2) =C xy.
\end{equation}
The map can be understood geometrically as a composition of two Manin-involutions $\gamma=\iota_q\circ \iota_p$, where one involution point, e.g. $p$, is a non-singular base point of the pencil, but the other involution point, $q=q(C)$, depends on the particular curve in the pencil, cf. \cite{KMQ}. Section \ref{conclusions} contains concluding remarks and sketches examples of quad-graphs on a strip beyond the $\ZZ^2$-lattice case.

\section{Boundary equations, folding and discrete boundary zero curvature conditions}\label{BC}
In this section, we first recall the notions of $3$D consistency and boundary consistency that are the integrability criteria for quad equations and their associated boundary equations respectively. % The following ideas are new.
Then, we introduce the notions of {\em dual boundary equation} and {\em dual boundary consistency condition}. We show that the dual boundary consistency condition gives rise to the {\em discrete boundary zero curvature condition}, which is the key structure in the open boundary reduction. This provides a systematic approach to deriving the {\em boundary matrices} appearing in the discrete boundary zero curvature condition. Explicit examples for the H1 and Q1($\delta=0$) equations from the ABS list are provided.
%\subsection{Boundary consistency}
\subsection{$3$D-consistent quad equations and discrete zero curvature conditions}\label{3D_cons}
%We first review the well-known integrable equations on quad-graphs without boundary, or the {\it bulk} equations.
Consider a quad equation, which is a partial difference equation defined on an elementary quadrilateral, as in  Figure~\ref{fig:quad},
\begin{equation}
  \label{eq:QQq1}
  Q(u,\widetilde{u},\widehat{u},\widehat{\widetilde{u}};\alpha,\beta)=0\,,
\end{equation}
where $u$ is a discrete field defined on the underlying quad-graph. We employ the ~$\widetilde{}~,~ \widehat{}$~  notations to denote forward shifts of $u$  along two independent directions,  and the lattice parameters $\alpha, \beta$ are  associated to the ~$\widetilde{}~,~ \widehat{}$~  directions respectively. For instance, if the underlying graph is the $\ZZ^2$-lattice,
\begin{equation}
\label{lattice}
 u=u(n,m)\,,\quad \widetilde{u}=u(n+1,m), \quad \widehat{u}=u(n,m+1)\,,\quad \cdots.
\end{equation}
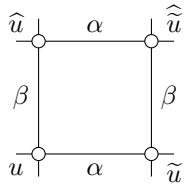
\begin{figure}[ht]
  \centering
  \begin{tikzpicture}[scale=1.5]
    \def\d{1}%
    \def\le{0.2}%
    \def\r{0.06}
    \def\ld{1.5}%
    \tikzstyle{nod1}= [circle, inner sep=0pt, fill=white, minimum size=5pt, draw]
    \tikzstyle{nod}= [circle, inner sep=0pt, fill=black, minimum size=5pt, draw]

    \coordinate (u00) at (-\le,0);
    \coordinate (u10) at (\le+\d,0);
    \coordinate (u01) at (-\le,\d);
    \coordinate (u11) at (\le+\d,\d);
    \coordinate (v00) at (0,-\le);
    \coordinate (v10) at (0,\d+\le);
    \coordinate (v01) at (\d,\le+\d);
    \coordinate (v11) at (\d,-\le);
    \draw[-] (u00)  node [below]{ $u$} -- node [below]{ $\alpha$} (u10) node [below]{ $\widetilde{u}$};
    \draw[-] (u01) node [above]{ $\widehat{u}$}-- node [above]{ $\alpha$} (u11)node [above]{ $\widehat{\widetilde{u}}$};
    \draw[-] (v00) -- node [left]{ $\beta$}(v10);
    \draw[-] (v01) -- node [right]{ $\beta$}(v11);

    \node[nod1] (u00) at (0,0) [label=below left: ] {};
    \node[nod1] (u10) at (\d,0) [label=below left: ] {};
    \node[nod1] (u01) at (0,\d) [label=below left:] {};
    \node[nod1] (u11) at (\d,\d) [label=below left:] {};

  %  \coordinate (w00) at (\le+\ld+\d,0);
  %  \coordinate (w10) at (\le+\ld+\d+\d,0);
  %  \coordinate (w01) at (\le+\ld+\d,\d);
%    \draw[-] (w00)  -- node [below]{ $\alpha$} (w10)  ;
%    \draw[thick] (w01) --  (w10);
%    \draw[-] (w00) -- node [left]{ $\alpha'$} (w01);
 %   \node[nod] (w10)at (\le+\ld+\d+\d,0) [label=below right: $x$] {};
 %   \node[nod] (w01) at (\le+\ld+\d,\d)[label=above left: $z$]{};
 %   \node[nod1] (w00) at (\le+\ld+\d,0) [label=below left:$y$] {};
  \end{tikzpicture}
\caption{Elementary quadrilateral supporting the bulk  equation.}
  \label{fig:quad}
\end{figure}

\begin{figure}[h]
	\centering
	\begin{tikzpicture}[scale=.55, decoration={markings,mark=at position 0.55 with {\arrow{latex}}}]
	\tikzstyle{nod1}= [circle, inner sep=0pt, fill=white, minimum size=5pt, draw]
	\tikzstyle{nod}= [circle, inner sep=0pt, fill=black, minimum size=5pt, draw]
	\def\lx{3}%
	\def\ly{1.22}%
	\def\lz{ (sqrt(\x*\x+\y*\y))}%
	\def\l{4}%
	\def\d{4}%
	\coordinate (u00) at (0,0);
	\coordinate (u10) at (\l,0);
	\coordinate (u01) at (\lx,\ly);
	\coordinate (u11) at (\l+\lx,\ly);
	\coordinate (v00) at (0,\d);
	\coordinate (v10) at (\l,\d);
	\coordinate (v01) at (\lx,\d+\ly);
	\coordinate (v11) at (\l+\lx,\d+\ly);
	\draw[-]  (u10)  % node[below] {$\widetilde{u}$}
	-- (u11) ;
	\draw [dashed]  (u01)  -- (u11);
	\draw [dashed]  (u00)--node [above]{$\beta$}(u01) ;
	\draw[-] (v00) --  (v10) -- (v11) -- (v01) -- (v00);
	\draw[-] (u11) --  (v11);
	\draw[-] (u00) -- node [left]{$\lambda$}(v00);
	\draw[-] (u10) -- (v10);
	\draw[dashed] (u01) -- (v01);
	\coordinate (u011) at (1.5*\lx,1.5* \ly);
	\draw[dashed] (u01) -- (u011) ;
	\draw[dashed] (u01) -- (.66*\lx,\ly) ;
	\draw[-] (u00) -- (-0.5*\lx,-0.5* \ly);
	\draw[-] (u00) -- (-.33*\lx,0);
	\draw[-] (u10) -- (0.33*\lx+\l,0);
	\draw[-] (u10) -- (-.5*\lx+\l,-.5*\ly);
	\draw[-] (u11) -- (1.33*\lx+\l,\ly);
	\draw[-] (u11) -- (-.5*\lx+\l,-.5*\ly);
	\draw[-]  (u11)-- (1.5*\lx+\l,1.5* \ly);
	\draw[-] (u00) % node[below] {$u$}
	-- node [below]{$\alpha$} (u10);
	\node[nod] (v00) at (0,\d) [label=above: $v$] {};
	\node[nod1] (v10) at (\l,\d) [label=above: $\widetilde{v}$] {};
	\node[nod1] (v01) at (\lx,\d+\ly) [label=above: $\widehat{v}$] {};
	\node[nod1] (v11) at (\l+\lx,\d+\ly) [label=above: $\widehat{\widetilde{v}}$] {};
	\node[nod] (u00) at (0,0) [label=below: $u$] {};
	\node[nod] (u10) at (\l,0) [label=below: $\widetilde{u}$] {};
	\node[nod] (u01) at  (\lx,\ly) [label=below: $\widehat{u}$] {};
	\node[nod1] (u11) at (\l+\lx,\ly) [label=below: $\widehat{\widetilde{u}}$] {};
	\end{tikzpicture}
	\caption{$3$D consistency: a quad equation can be imposed on the six faces of a cube. The black dots indicate the initial values. %The three ways of computing $\widehat{\widetilde{v}}$ from those values give the same result.
} \label{fig:31}
\end{figure}
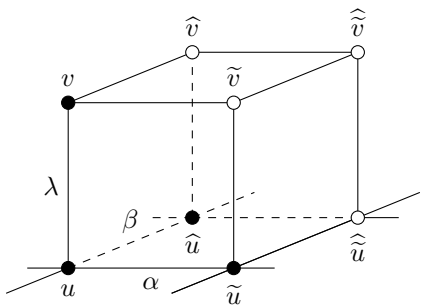

The {\em bulk dynamics} on a quad-graph is determined by $Q=0$ subject to well-posed initial data \cite{AV,VdK1}. A {$3$D consistency} condition  was introduced as a defining criterion for the integrability of quad equations \cite{BS, nijhoff2002lax, NW1}. Namely, a quad equation \eqref{eq:QQq1} is said to be integrable if it can be consistently defined on a cube, see  Figure~\ref{fig:31}. Given the initial values  $u, \widetilde{u}, \widehat{u}, v$ on a cube, the three possible ways to compute $\widehat{\widetilde{v}}$ must give the same result.

A classification of scalar $3$D-consistent equations, known as the ABS list \cite{adler2003classification}, was obtained under the following assumptions:
\begin{enumerate}%[label=(\roman*)]
  \item  ${\cal D}_4$-symmetry, \ie
\begin{equation}
  \label{eq:D4}
    Q(u,\widetilde{u},\widehat{u},\widehat{\widetilde{u}};\alpha,\beta)=  \omega\, Q(\widetilde{u}, u,\widehat{\widetilde{u}},\widehat{u};\alpha,\beta) = \delta\,Q(u,\widehat{u},\widetilde{u}, \widehat{\widetilde{u}} ;\beta,\alpha)\,,\quad \omega= \pm 1, \quad \delta = \pm 1\,.
  \end{equation}
  The case where $\omega=-1$ is excluded, cf.~\cite{adler2003classification}.
\item Affine-linearity with respect to each of its fields.
\item The so-called tetrahedron property, \ie $\widehat{\widetilde{v}}$ does not depend on $u$, but only on $v$, $\widetilde{u}$ and $\widehat{u}$ (and lattice parameters).
\end{enumerate}

$3$D consistency, together with the above properties,
allows one to derive discrete zero curvature conditions \cite{BS,nijhoff2002lax}. It follows from $Q(u,\widetilde{u},v,\widetilde{v},\alpha,\lambda)=0$ and the affine-linearity of $Q$ that $\widetilde{v}$ can be expressed as a M\"obius transformation acting on $v$:
\begin{equation}
\label{eq:MOB1}
\widetilde{v} = \frac{l_1 \,v+ l_2}{l_3 \,v+l_4} =  L \,[v]  \,, \quad L=L(\widetilde{u},u;\alpha,\lambda)=  \mu L_c
\end{equation}
where $\mu$ is a scalar function and the `core' of the matrix $L$ is $L_c=\bma l_1 &  l_2 \\ l_3 & l_4\ema$.
Similarly, one has $\widehat{v} = M\,[v]$, $M=L(\widehat{u},u;\beta,\lambda)=\nu M_c$.
By composition of M\"obius transformations, due to the $3$D consistency, we get the discrete zero curvature condition
\begin{equation}
  \label{eq:3mllm1}
           L(\widehat{\widetilde{u}},\widetilde{u};\beta,\lambda) \,L(\widetilde{u},u;\alpha,\lambda)\,[v] = L(\widehat{\widetilde{u}},\widehat{u};\alpha,\lambda)\, L(\widehat{u},u;\beta,\lambda)\,[v],
\end{equation}
which holds when $Q(u,\widetilde{u}, \widehat{u}, \widehat{\widetilde{u}};\alpha,\beta)=0$. Note that since the lattice parameter $\alpha$ does not depend on the hat-shift and the lattice parameter $\beta$ does not depend on the tilde-shift, \eqref{eq:3mllm1}  can be written as
%\begin{equation} \label{zcc}
$\widetilde{M} \,L \, [v] = \widehat{L}\, M\, [v]$.
%\end{equation}
Note that this zero curvature condition is formulated as a projective identity. The scalars $\mu,\nu$ in the Lax matrices $L, M$ are irrelevant in the action of the M\"obius transformations. In order to obtain a true matrix equation
\begin{equation} \label{zcc}
\widetilde{M} \,L = \widehat{L}\, M,
\end{equation}
which holds when \eqref{eq:QQq1} is satisfied, one has to fix the scalars $\mu,\nu$ appropriately. This is explained in detail in \cite{BHQK} for several classes of 3D consistent equations. One option is to choose the normalisation so that the determinants of the Lax matrices equal $1$. However, this may introduce unnecessary (square) roots to deal with. For this reason we allow more freedom. For equations from the ABS list, the normalization can be fixed such that
%\begin{lemma}
\begin{equation}
  \label{eq:invL}
\det L(u,\widetilde{u};\alpha,\lambda)=\ell(\alpha,\lambda)\quad L(u,\widetilde{u};\alpha,\lambda) L(\widetilde{u},u;\alpha,\lambda) = \pm \ell(\alpha,\lambda)\id \,,
\end{equation}
where  $\ell$ is a function depending only on the parameters. With appropriate normalisation, one can introduce an auxiliary vector $\Psi$ on the elementary quadrilateral and interpret (\ref{zcc}) as the compatibility condition, ${\widehat{\widetilde{\Psi}}}=\widehat{\widetilde{{\Psi}}}$, of the discrete linear Lax system
\begin{equation}
\widetilde{\Psi}=L\Psi, \qquad \widehat{\Psi}=M\Psi\,.
\end{equation}

\subsection{Boundary consistency and integrable boundary equations}
\begin{figure}[ht]
	\centering
	\begin{tikzpicture}[scale=1.8]
	\def\d{1}%
	\def\le{0.2}%
	\def\r{0.06}
	\def\ld{1.5}%
	\tikzstyle{nod1}= [circle, inner sep=0pt, fill=white, minimum size=5pt, draw]
	\tikzstyle{nod}= [circle, inner sep=0pt, fill=black, minimum size=5pt, draw]
	
	\coordinate (u00) at (-\le,0);
	\coordinate (u10) at (\le+\d,0);
	\coordinate (u01) at (-\le,\d);
	\coordinate (u11) at (\le+\d,\d);
	\coordinate (v00) at (0,-\le);
	\coordinate (v10) at (0,\d+\le);
	\coordinate (v01) at (\d,\le+\d);
	\coordinate (v11) at (\d,-\le);
	%\draw[-] (u00)  node [below]{ $u$} -- node [below]{ $\alpha$} (u10) node [below]{ $\widetilde{u}$};
	%\draw[-] (u01) node [above]{ $\widehat{u}$}-- node [above]{ $\alpha$} (u11)node [above]{ $\widehat{\widetilde{u}}$};
	%\draw[-] (v00) -- node [left]{ $\beta$}(v10);
	%\draw[-] (v01) -- node [right]{ $\beta$}(v11);

	%\node[nod1] (u00) at (0,0) [label=below left: ] {};
	%\node[nod] (u10) at (\d,0) [label=below left: ] {};
	%\node[nod] (u01) at (0,\d) [label=below left:] {};
	%\node[nod1] (u11) at (\d,\d) [label=below left:] {};
	
	\coordinate (w00) at (\le+\ld+\d,0);
	\coordinate (w10) at (\le+\ld+\d+\d,0);
	\coordinate (w11) at (\le+\ld+\d+\d,\d);
	\draw[thick] (w00)  -- node [below]{ $\alpha$} (w10)  ;
	\draw[line width=1.5pt] (w00) --  (w11);
	\draw[thick] (w10) -- node [right]{ $\beta$} (w11);
	\node[nod1] (w10)at (\le+\ld+\d+\d,0) [label=below right: $y$] {};
	\node[nod1] (w11) at (\le+\ld+\d+\d,\d)[label=above left: $z$]{};
	\node[nod1] (w00) at (\le+\ld+\d,0) [label=below left:$x$] {};
	\end{tikzpicture}
	\caption{Elementary triangle supporting a boundary equation. The thick line represents the boundary connecting the boundary fields.}
	\label{fig:tri}
\end{figure}
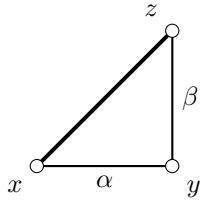

Inspired by \cite{M1}, it was shown it \cite{CCZ} that by dualizing a cellular decomposition of a surface with boundary, one naturally obtains a quad-graph where triangular faces represent the boundary. A {\bf boundary equation} of the form
\begin{equation}
\label{eq:lq}
q(x, y,z; \alpha, \beta)=0\,,
\end{equation}
is required to hold on each elementary triangle, as in Figure~\ref{fig:tri}, similarly to \eqref{eq:QQq1} holding on a quadrilateral.
By convention, the first and third arguments in  $q$, \ie $x,z$ in \eqref{eq:lq},  are the values of the field at the boundary vertices; the second argument, \ie $y$ in \eqref{eq:lq}, is a bulk value of the field. The parameters  $\alpha, \beta$, presumably independent of each other, are the lattice parameters associated to edges connecting the bulk value $y$ to the boundary values $x$ and $z$ respectively.

It is natural to require boundary equations $q(x,y,z;\alpha,\beta)=0$ to possess the following properties (i)-(iii):
\begin{enumerate}
\item[(i)] {\bf Affine-linearity}: $q$ is affine-linear with respect to the boundary fields:
    \begin{equation}\label{eq:generalform}
    q(x,y,z;\alpha,\beta) = m_1xz+m_2x+m_3z+m_4=q_1z+q_2=q_3x+q_4\,.
  \end{equation}
  where $m_i=m_i(y;\alpha,\beta)$, $q_i=q_i(x,y;\alpha,\beta)$ when $i=1,2$ and $q_i=q_i(z,y;\alpha,\beta)$ when $i=3,4$.
\item[(ii)] {\bf Nondegeneracy}: $q_{xz}q-q_xq_z\neq 0$. This amount to $
    m_1m_4-m_2m_3\neq 0 $, and ensures that $q=0$ can be solved to express $x$ (resp. $z$) in terms of $z$ (resp. $x$) and (possibly) $y$.
  \begin{equation}\label{eq:mob1}
    q(x,y,z;\alpha,\beta) = 0 \quad\Rightarrow\quad x=-\frac{q_4}{q_3}~~\text{or}~~ z = -\frac{q_2}{q_1}\,.
  \end{equation}
\item[(iii)] {\bf $\ZZ_2$-symmetry}: there exists a function $h=h(\alpha,\beta)$ such that $q(x,y,z;\alpha,\beta)    =h(\alpha,\beta)  q(z,y,x;\beta,\alpha)$. For consistency, we have $h(\alpha,\beta)h(\beta,\alpha)=1$.
\end{enumerate}

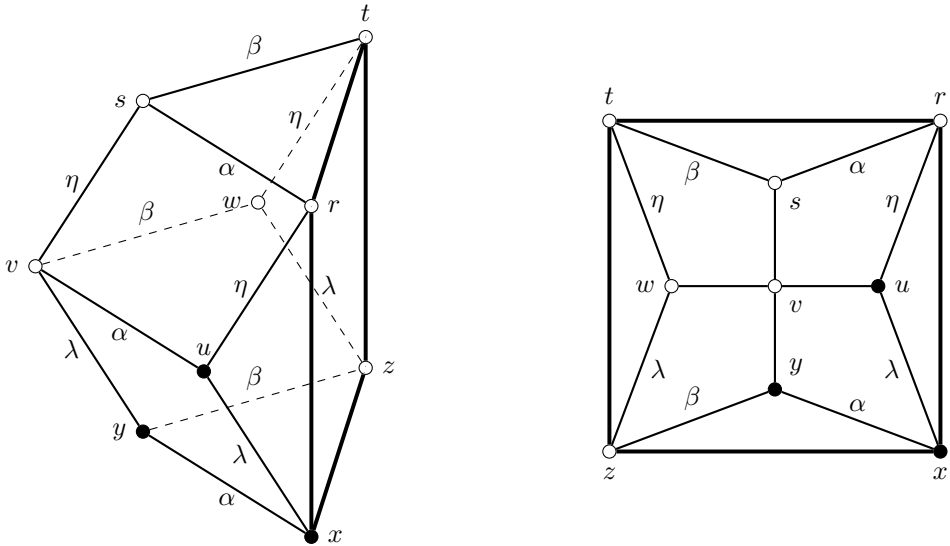
\begin{figure}[h]
  \centering
  \begin{subfigure}[b]{0.4\textwidth}
  \begin{tikzpicture}[scale=2.2]
    \tikzstyle{nod1}= [circle, inner sep=0pt, minimum size=5pt, draw]
    \tikzstyle{nod}= [circle, inner sep=0pt, fill=black, minimum size=5pt, draw]
    \tikzstyle{vertex}=[circle,minimum size=20pt,inner sep=0pt]
    \tikzstyle{selected vertex} = [vertex, fill=red!24]
    \tikzstyle{selected edge} = [draw,line width=1.5pt,-]
    \tikzstyle{edge} = [draw, thick,-,black] %inner lines
    \tikzstyle{ddedge} = [draw, densely dotted,-,black] %dotted lines
    \tikzstyle{dedge} = [draw, dashed,-,black] %dashed lines
    \tikzstyle{eedge} = [draw,line width=1.5pt,-,black] %outer lines

    \pgfmathsetmacro \ll {6}% inner lenght
    \pgfmathsetmacro \rl {7}% inner lenght
    \pgfmathsetmacro \h {2}% inner lenght
    \pgfmathsetmacro \an {32}%projection angle a
    \pgfmathsetmacro \bn {16}%projection angle b
    \pgfmathsetmacro \cn {18}%projection angle c
    \pgfmathsetmacro \ra {.2}%projection ration
    \pgfmathsetmacro \s {.4}%separationn distance

    \pgfmathsetmacro \llx {{\ra*\ll*cos(\an)}}
    \pgfmathsetmacro \lly {{\ra*\ll*sin(\an)}}
    \pgfmathsetmacro \rlx {{\ra*\rl*cos(\bn)}}
    \pgfmathsetmacro \rly {{\ra*\rl*sin(\bn)}}
    \pgfmathsetmacro \dh {{\h*tan(\cn)}}
    % 1st floor v, v1, v4, v14
    \node[nod] (v) at (0,0) [label=right:$x$] {};
    \node[nod] (v1) at (-\llx,\lly) [label=left:$y$] {};
    \node[nod1] (v14) at (\rlx-\llx,\rly+\lly) [label=right:$z$] {};
    \draw[dedge] (v1) --node [above] {$\beta$} (v14)  ;
    \draw[edge] (v1) -- node [below] {$ \alpha$} (v) ;
    \draw[eedge] (v) --  (v14)  ;
    % 3rd floor w, w1, w4, w14
    \node[nod1] (w) at (0,\h) [label=right: $r$] {};
    \node[nod1] (w1) at (-\llx,\lly+\h) [label=left: $s$] {};
    \node[nod1] (w14) at (\rlx-\llx,\rly+\lly+\h) [label=above:$t$] {};
    \draw[edge] (w1) -- node [above]{$\beta$}(w14);
    \draw[edge] (w1) -- node [below]{$\alpha$}(w) ;
    \draw[eedge] (w) -- (w14) ;
    %2nd floor u1,u2,u3,u4,u5,u6
    \node[nod] (u1) at (-\dh,\h/2) [label=above:$u$] {};
    \draw[edge] (v) --  node [left] {$\lambda$} (u1) -- node [left] {$\eta$} (w);
    \draw[eedge] (v) -- (w);
    \node[nod1] (u5) at (-\dh -\llx,\h/2+\lly) [label=left:$v$] {};
    \node[nod1] (u6) at (-\dh+\rlx-\llx,\rly+\lly+\h/2) [label= left :$w$] {};
    \draw[edge] (u1) -- node [below]{$\alpha$} (u5);
    \draw[dedge] (u5) --node [above]{$\beta$} (u6);
    \draw[edge] (v1) -- node [left]{$\lambda$}   (u5) --  node [left]{$\eta$} (w1);
    \draw[dedge] (v14) --  node [right]{$\lambda$}  (u6) -- node [left]{$\eta$}  (w14) ;
    \draw[eedge] (v14)  --  (w14) ;
    %shadow projection
    \draw [dotted, opacity =.4] (0,0) --(0,\h) -- (\rlx-\llx,\rly+\lly+\h)                  --  (\rlx-\llx,\rly+\lly) --cycle;
  \end{tikzpicture}
\end{subfigure}
\begin{subfigure}[b]{0.4\textwidth}
  \begin{tikzpicture}[scale=1.1]
    \tikzstyle{nod1}= [circle, inner sep=0pt, minimum size=5pt, draw]
    \tikzstyle{nod}= [circle, inner sep=0pt, fill=black, minimum size=5pt, draw]
    \tikzstyle{vertex}=[circle,minimum size=20pt,inner sep=0pt]
    \tikzstyle{selected vertex} = [vertex, fill=red!24]
    \tikzstyle{selected edge} = [draw,line width=1.5pt,-]
    \tikzstyle{edge} = [draw, thick,-,black] %inner lines
    \tikzstyle{ddedge} = [draw, densely dotted,-,black] %dotted lines
    \tikzstyle{dedge} = [draw, dashed,-,black] %dashed lines
    \tikzstyle{eedge} = [draw, line width=1.5pt,-,black] %outer lines
    \pgfmathsetmacro \lv {1.1}% inner lenght
    \pgfmathsetmacro \lh {0.8}% inner lenght
    \pgfmathsetmacro \ld {2}% inner lenght
    \pgfmathsetmacro \an {38}%projection angle a

    \pgfmathsetmacro \llx {{\ld*tan(\an)}}
    \pgfmathsetmacro \lly {{\ld/sin(\an)}}
    % 1st floor v, v1, v4, v14
    \node (em) at (0,0) [label=right:${}$] {};
    \node[nod1] (z) at (0+\lh,\lv) [label=below:$z$] {};
    \node[nod] (x) at (2*\ld+\lh,0+\lv) [label=below:$x$] {};
    \node[nod1] (bz) at (0+\lh,2*\ld+\lv) [label=above:$t$] {};
    \node[nod1] (bx) at (2*\ld+\lh,2*\ld+\lv) [label=above:$r$] {};
    \draw[eedge] (z) --  (x)  -- (bx) -- (bz) -- (z);

    \node[nod1] (by) at (0+\lh+\ld,\lv+\lly) [label=below right:$s$] {};
    \node[nod1] (w) at (2*\ld+\lh-\lly,0+\lv+\ld) [label=left:$w$] {};
    \node[nod] (u) at (0+\lh+\lly,\lv+\ld) [label=right:$u$] {};
    \node[nod] (y) at (0+\lh+\ld,2*\ld+\lv-\lly) [label=above right :${y}$] {};
    \draw[edge] (bz) -- node [below] {$\beta$} (by)  -- node [below] {$\alpha$}(bx) -- node [left] {$\eta$} (u) --node [left] {$\lambda$} (x) --node [above] {$\alpha$} (y) --node [above] {$\beta$} (z) -- node [right] {$\lambda$} (w) -- node [right] {$\eta$}(bz);
    \node[nod1] (v) at (0+\lh+\ld,\lv+\ld) [label=below right:$v$] {};
    \draw[edge] (w) --  % node [below] {$\beta$}
    (v) --  % node [below] {$\alpha$}
    (u);
    \draw[edge] (by) --  % node [right] {$\eta$}
    (v) -- % node [right] {$\lambda$}
    (y);
  \end{tikzpicture}
\end{subfigure}
  \caption{Boundary consistency around half of a rhombic dodecahedron (left) and its planar projection (right), where $Q=0$ is imposed on  $4$ quadrilaterals and $q=0$ is imposed on $4$ triangles. }
    \label{fig:bccy}
\end{figure}

In \cite{CCZ}, a criterion, called {\em boundary consistency}, was introduced to select special boundary equations $q=0$ yielding  {\em integrable boundary conditions} for an integrable quad equation $Q=0$. The criterion is defined as follows.
\begin{definition}
\label{Def_integ_bd}
A boundary equation $q=0$ is {\bf boundary consistent} with an integrable quad equation $Q=0$ if there is an involutive function $\sigma$ between the parameters, $\beta=\sigma(\alpha)$ and $\eta=\sigma(\lambda)$, such that the initial value problem on the half rhombic dodecahedron in Figure~\ref{fig:bccy} is well-posed, \ie if the three ways of computing $t$
from initial values $x, y, u$ yield the same result. A boundary equation which is boundary consistent is called {\bf integrable}.
\end{definition}

A list of integrable boundary equations for $Q$ belonging to the ABS list was obtained in \cite{CCZ}, using the idea of ``folding the square into a triangle''. In a nutshell, the idea is to look for $q$ in the form of $Q$ with one of the corner eliminated in favour of two others (square into triangle) via some unknown function. The ansatz is then inserted into the equations for boundary consistency of Figure~\ref{fig:bccy} in order to find the unknown function.  More explicitly, one considers
\begin{equation}
\label{eq:k}
Q(x,y,k(x,y,\alpha),z;\alpha,\sigma(\alpha))\,,
\end{equation}
and looks for a function $k$ (linear fractional in $y$) and an involution $\sigma$ such that:
\begin{enumerate}
	\item[(a)] The following factorization holds
	\begin{equation} Q(x,y,k(x,y,\alpha),z;\alpha,\sigma(\alpha))=f(x,y,\alpha)q(x,y,z;\alpha)\,,
	\end{equation}
	with $q$ having properties (i)-(iii) above;
	\item[(b)] The boundary consistency condition of Figure~\ref{fig:bccy} hold.
\end{enumerate}

The function $k$ was a useful tool in \cite{CCZ} but its role was not fully exploited. In the next section, we elucidate the role of $k$ as originating from a dual boundary equation associated to $q$. We also separate more clearly step (a), which contains no information about integrability but simply provides potential candidates for integrable boundary equations, and step (b) which deals with finding integrable boundary equations. The notion of dual boundary equation will enable us to formulate a novel boundary consistency condition we call (discrete) boundary zero curvature condition.

\subsection{Folding and dual boundary equations}\label{folding}
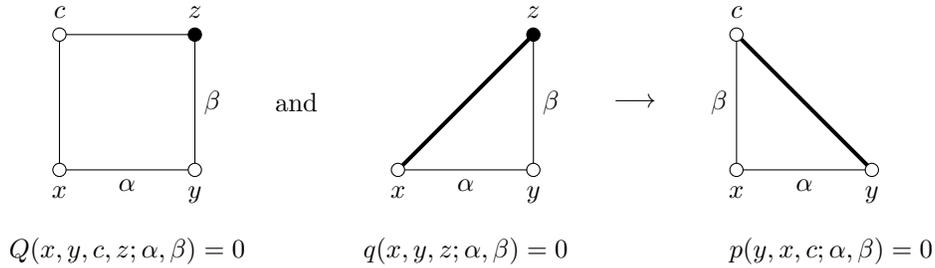
\begin{figure}[htb]
	\begin{center}
		\begin{tikzpicture}[scale=1.8]
		\def\l{1}%
		\def\d{1.5}%
		\def\le{.1}
		\def\b{.3}
		\tikzstyle{nod1}= [circle, inner sep=0pt,fill=white,  minimum size=5pt, draw]
		\tikzstyle{nod}= [circle, inner sep=0pt, fill=black, minimum size=5pt, draw]
		% left triangle
		\coordinate (w00) at (0,0);
		\coordinate (w10) at (\l,0);
		\coordinate (w01) at (0,\l);
		\coordinate (w11) at (\l,\l);
		% middle square
		\coordinate (u00) at (\l+\d,0);
		\coordinate (u10) at (\l+\l+\d,0);
		\coordinate (u01) at (\l+\d,\l);
		\coordinate (u11) at (\l+\l+\d,\l);
		% right triangle
		\coordinate (v00) at (\l+\l+\d+\d,0);
		\coordinate (v10) at (\l+\l+\d+\d+\l,0);
		\coordinate (v01) at (\l+\l+\d+\d,\l);
		\coordinate (v11) at (\l+\l+\d+\d+\l,\l);
		
		\draw[-] (w00) --   node [below]{ $\alpha$}  (w10) --node [right]{ $\beta$} (w11) -- (w01) -- (w00) ;%left square
		%    \draw[dashed] (w10) --  (w01);%(zx) diag middle square
		%    \draw[dashed] (w00) --   (w11);%(yc) diag middle square
		\node[nod1](w00) at (0,0)  [label=below: $x$] {};
		\node[nod1](w10) at (\l,0) [label=below: $y$] {};
		\node[nod1](w01) at (0,\l) [label=above: $c$ ] {};
		\node[nod](w11) at (\l,\l) [label=above: $z$ ] {};
		\node (q1) at (.5*\l,-2*\b) {$Q(x,y,c,z;\alpha,\beta)=0$};
		
		\draw[-] (u00) --  node [below]{ $\alpha$} (u10) -- node [right]{ $\beta$} (u11);%middle triangle
		\draw[line width=1.5pt] (u00) --  (u11);
		\node[nod1]  (u00) at (\l+\d,0) [label=below: $x$] {};
		\node[nod1] (u10) at (\l+\l+\d,0) [label=below: $y$] {};
		\node[nod] (u11) at (\l+\l+\d,\l) [label=above: $z$ ] {};
		\node (Q) at (\l+\d+.5*\l,-2*\b) {$q(x,y,z;\alpha, \beta)=0$};

			\draw[-] (v10) --  node [below]{ $\alpha$} (v00)  -- node [left]{ $\beta$} (v01) ;%right triangle
		\draw[line width=1.5pt] (v10) --  (v01);    %diag in right triangle
		\node[nod1] (v10) at (\l+\l+\d+\d+\l,0)[label=below:$y$ ] {};
		\node[nod1] (v00) at (\l+\l+\d+\d,0) [label=below:$x$] {};
		\node[nod1] (v01) at (\l+\d+\d+\l,\l) [label=above:$c$] {};
		%    \node[nod1] (v01) at (\l+\d+\d+\l,\l) [label=above: $z$ ] {};
		%   \node (q22) at (\l+\l+\d+\d+.7*\l,\l+1.5*\b) {$p(c,z,y;\alpha, \beta)=0$};
		
		%	\node (1) at (\l+.5*\d,.5*\l) {$ \Leftrightarrow  $};
		\node (2) at (\l+.5*\d+\l+\d,.5*\l) {$\longrightarrow  $};
		\node (q2) at (\l+\l+\d+\d+.7*\l,-2*\b) {$p(y,x,c;\alpha, \beta)=0$};
		
		\node (1) at (\l+.5*\d,.5*\l) {and};
%		\node (2) at (\l+.3*\d+\l+\d,.5*\l) {$\longrightarrow  $};
		%    \node (q2) at (\l+\l+\d+\d+.7*\l,-2*\b) {$q(x,y,z;\alpha, \beta)=0$};
		
		\end{tikzpicture}
	\end{center}
	\caption{\label{Folding_fig1} Folding procedure: obtaining $p$ from $q$ and $Q$. We use $q=0$ to eliminate $z$ in $Q$ (this elimination is shown by the black dots). $\beta$ is independent of $\alpha$ at this stage, it will become $\sigma(\alpha)$ when considering integrability.}
\end{figure}

We formalise the idea of folding as follows. Let an integrable quad equation $Q=0$ be given. We will suppose a boundary equation $q=0$, where $q$ satisfies properties (i)-(iii), is given such that if we use $q(x,y,z;\alpha,\beta)=0$ to express $z$ as in \eqref{eq:mob1} and eliminate $z$ in $Q(x,y,c,z;\alpha,\beta)$, there exist a polynomial function $\chi(x,y; \alpha,\beta)$ and a polynomial $p(y,x,c;\alpha,\beta)$ {\it satisfying properties (i)-(iii)}, with the following relation holding
\begin{equation}\label{p}
q_1(x,y;\alpha,\beta)Q(x,y,c,  -\frac{q_2(x,y;\alpha,\beta)}{q_1(x,y;\alpha,\beta)};\alpha,\beta)=  \chi(x,y; \alpha,\beta)\,   p(y,x,c;\alpha,\beta)\,.
\end{equation}
The situation is illustrated in Figure~\ref{Folding_fig1}.

We need to address to what extent the folding procedure provides a map from $q$ to $p$. Let us note that if the factorization \eqref{p} exists then $p$ and $\chi$ are unique up to an overall function of the parameters $\alpha,\beta$ only.  Therefore, strictly speaking to each $q$ we associate an equivalence class $[p]$ of boundary equations defined by the relation $p\sim p^\ast$ if and only $p(y,x,c;\alpha, \beta)=g(\alpha,\beta)\,p^\ast(y,x,c;\alpha, \beta)$ for some function $g(\alpha,\beta)$\footnote{Note that the corresponding relation on $\chi$, $\chi^\ast$ is $\chi^\ast(x,y;\alpha,\beta)=g(\alpha,\beta)\chi(x,y;\alpha,\beta)$.}. Of course, as far as the boundary equation $p=0$ is concerned any representative $p$ in $[p]$ yields the same relation on $y,x,c$. In particular, the same holds true for $q=0$ so it is more appropriate to think of the folding as mapping an equivalence class $[q]$ to an equivalence class $[p]$. In practice, we can use any representative we like. In the rest of the paper, we will simply use the notation $q$ and $p$ as it should not lead to confusion whether we mean a representative or the class. Also, we will omit multipliers and dependence on variables/parameters when these are clear from the context, e.g. equation (\ref{p}) can be shortly written as $
q_1Q(x,y,c,-\frac{q_2}{q_1};\alpha,\beta)\propto p(y,x,c;\alpha,\beta)$.

\begin{figure}[htb]
	\begin{center}
		\begin{tikzpicture}[scale=1.8]
		\def\l{1}%
		\def\d{1.5}%
		\def\le{.1}
		\def\b{.3}
		\tikzstyle{nod1}= [circle, inner sep=0pt,fill=white,  minimum size=5pt, draw]
		\tikzstyle{nod}= [circle, inner sep=0pt, fill=black, minimum size=5pt, draw]
		% left triangle
		\coordinate (w00) at (0,0);
		\coordinate (w10) at (\l,0);
		\coordinate (w01) at (0,\l);
		\coordinate (w11) at (\l,\l);
		% middle square
		\coordinate (u00) at (\l+\d,0);
		\coordinate (u10) at (\l+\l+\d,0);
		\coordinate (u01) at (\l+\d,\l);
		\coordinate (u11) at (\l+\l+\d,\l);
		% right triangle
		\coordinate (v00) at (\l+\l+\d+\d,0);
		\coordinate (v10) at (\l+\l+\d+\d+\l,0);
		\coordinate (v01) at (\l+\l+\d+\d,\l);
		\coordinate (v11) at (\l+\l+\d+\d+\l,\l);
		
		\draw[-] (w00) --   node [below]{ $\alpha$}  (w10) --node [right]{ $\beta$} (w11) -- (w01) -- (w00) ;%left square
		%    \draw[dashed] (w10) --  (w01);%(zx) diag middle square
		%    \draw[dashed] (w00) --   (w11);%(yc) diag middle square
		\node[nod](w00) at (0,0)  [label=below: $x$] {};
		\node[nod1](w10) at (\l,0) [label=below: $y$] {};
		\node[nod1](w01) at (0,\l) [label=above: $c$ ] {};
		\node[nod1](w11) at (\l,\l) [label=above: $z$ ] {};
		\node (q1) at (.5*\l,-2*\b) {$Q(x,y,c,z;\alpha,\beta)=0$};
		
		\draw[-] (u00) --  node [below]{ $\alpha$} (u10) -- node [right]{ $\beta$} (u11);%middle triangle
		\draw[line width=1.5pt] (u00) --  (u11);
		\node[nod]  (u00) at (\l+\d,0) [label=below: $x$] {};
		\node[nod1] (u10) at (\l+\l+\d,0) [label=below: $y$] {};
		\node[nod1] (u11) at (\l+\l+\d,\l) [label=above: $z$ ] {};
		\node (Q) at (\l+\d+.5*\l,-2*\b) {$q(x,y,z;\alpha, \beta)=0$};

			\draw[-] (v10) --  node [right]{ $\beta$} (v11)  -- node [above]{ $\alpha$} (v01) ;%right triangle
		\draw[line width=1.5pt] (v10) --  (v01);    %diag in right triangle
		\node[nod1] (v10) at (\l+\l+\d+\d+\l,0)[label=below:$y$ ] {};
		\node[nod1] (v11) at (\l+\l+\d+\d+\l,\l) [label=above:$z$] {};
		\node[nod1] (v01) at (\l+\d+\d+\l,\l) [label=above:$c$] {};
		%    \node[nod1] (v01) at (\l+\d+\d+\l,\l) [label=above: $z$ ] {};
		%   \node (q22) at (\l+\l+\d+\d+.7*\l,\l+1.5*\b) {$p(c,z,y;\alpha, \beta)=0$};
		
		%	\node (1) at (\l+.5*\d,.5*\l) {$ \Leftrightarrow  $};
		\node (2) at (\l+.5*\d+\l+\d,.5*\l) {$\longrightarrow  $};
		\node (q2) at (\l+\l+\d+\d+.7*\l,-2*\b) {$p(c,z,y;\alpha, \beta)=0$};
		
		\node (1) at (\l+.5*\d,.5*\l) {and};
%		\node (2) at (\l+.3*\d+\l+\d,.5*\l) {$\longrightarrow  $};
		%    \node (q2) at (\l+\l+\d+\d+.7*\l,-2*\b) {$q(x,y,z;\alpha, \beta)=0$};
		
		\end{tikzpicture}
	\end{center}
	\caption{\label{Folding_fig2} Folding procedure: obtaining $p$ from $q$ and $Q$ by eliminating $x$ instead of $z$.}
\end{figure}
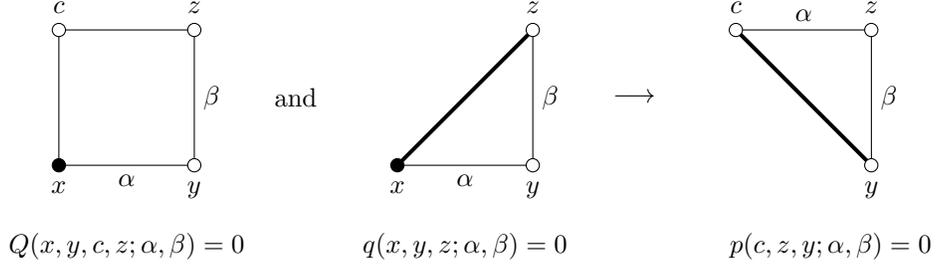

If the above folding occurs, then alternatively we could decide to use $q(x,y,z;\alpha,\beta)=0$ to eliminate $x$ instead of $z$. This will give rise to {\em the same} boundary equation $p=0$, see Figure~\ref{Folding_fig2}.
Moreover, if one would use the newly obtained boundary equation $p=0$ to eliminate either $c$ or $y$, one will get back the original boundary equation $q=0$.

\begin{lemma}[Duality] \label{lem:folding}
Let $Q=Q(x,y,c,z;\alpha,\beta)$ be a multi-linear function with ${\cal D}_4$ symmetry, and let the boundary equation (\ref{eq:generalform}) possess properties (i)-(iii).
Suppose that $p=p(y,x,c;\alpha,\beta)$ is a divisor of $q_1Q(x,y,c,  -\frac{q_2}{q_1};\alpha,\beta)$ and that $p$ also possesses properties (i)-(iii), so that $p$ can be written variously as
\begin{equation}
p(y,x,c;\alpha,\beta)=p_1(y,x;\alpha,\beta)c+p_2(y,x;\alpha,\beta)=p_3(c,x;\alpha,\beta)y+p_4(c,x;\alpha,\beta).
\end{equation}
Then, we have: $
\begin{aligned}[t]
\text{a)}\quad &q_3Q(-\frac{q_4}{q_3},y,c,z;\alpha,\beta)\propto p(c,z,y;\alpha,\beta),\\
\text{b)}\quad &p_1Q(x,y,-\frac{p_2}{p_1},z;\alpha,\beta)\propto q(x,y,z;\alpha,\beta),\\
\text{c)}\quad &p_3Q(x,-\frac{p_4}{p_3},c,z;\alpha,\beta)\propto q(z,c,x;\alpha,\beta).
\end{aligned}$
\end{lemma}
\noindent
\prf
\begin{enumerate}
\item[\text{a)}] Because $q$ has $\ZZ_2$ symmetry, apart from $x=-\frac{q_4(z,y;\alpha,\beta)}{q_3(z,y;\alpha,\beta)}$, we also have
$x=-\frac{q_2(z,y;\beta,\alpha)}{q_1(z,y;\beta,\alpha)}$. Due to ${\cal D}_4$ symmetry we have
\begin{equation}
q_3(z,y;\alpha,\beta)Q(-\frac{q_4(z,y;\alpha,\beta)}{q_3(z,y;\alpha,\beta)},y,c,z;\alpha,\beta)
=h(\alpha,\beta)q_1(z,y;\beta,\alpha)\delta Q(z,y,c,-\frac{q_2(z,y;\beta,\alpha)}{q_1(z,y;\beta,\alpha)};\beta,\alpha)\,,
\end{equation}
which admits the divisor $p(y,z,c;\beta,\alpha)\propto p(c,z,y;\alpha,\beta)$.
\item[\text{b)}]
Since $Q$ is a multivariate affine-linear polynomial, we write for convenience
$
Q(x,y,c,z;\alpha,\beta) = Q_1  c z +Q_2c +Q_3 z+Q_4
$,
where $Q_j=Q_j(x,y;\alpha,\beta)$, $j=1,\dots,4$ are multivariate affine-linear polynomials in $x,y$. Substitution of $z=-\frac{q_2}{q_1}$ and multiplying by $q_1$ gives
\begin{equation}
\label{eq_p1}
-(Q_1  c+Q_3)q_2  +(Q_2c+Q_4)q_1
\end{equation}
which vanishes when $p(y,x,c;\alpha,\beta)=p_1c+p_2=0$. Setting $p=0$, expressing $c$ in terms of $x,y$ and substituting in $Q$ yields
\begin{equation}
p_1Q(x,y,-\frac{p_2}{p_1},z,\alpha,\beta)=-(Q_1  z+Q_2)p_2  +(Q_3z+Q_4)p_1\,,
\end{equation}
which is a multivariate polynomial in $x,y,z$ and is linear in $z$. This polynomial vanishes for $z=-\frac{q_2}{q_1}$ in view of \eqref{eq_p1} so it must be proportional to $q_1 z+q_2=q(x,y,z;\alpha,\beta)$.
\item[\text{c)}] This follows from b) using the ${\cal D}_4$ symmetry of $Q$ and the $\ZZ_2$ symmetry of $q,p$. \finprf
\end{enumerate}

Lemma \ref{lem:folding} leads us to the following definition.
\begin{definition} \label{def_dual}
If $Q,q,p$ satisfy the conditions in Lemma \ref{lem:folding}, then we say that $p=0$ is the {\bf dual boundary equation} of $q=0$ and vice versa. The quad equation $Q=0$ is said to have a pair of dual boundary equations $p=0$, $q=0$.
\end{definition}

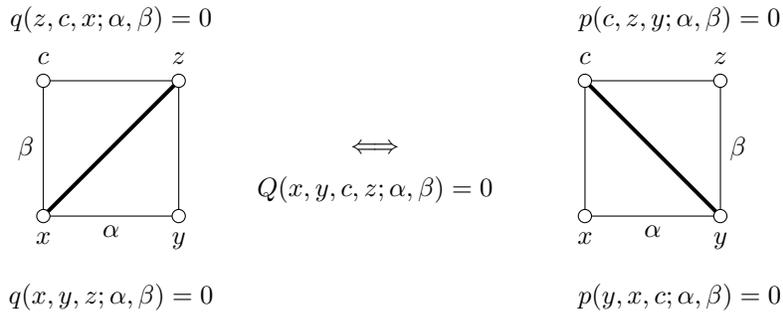
\begin{figure}[htb]
	\begin{center}
		\begin{tikzpicture}[scale=1.8]
		\def\l{1}%
		\def\d{1.5}%
		\def\le{.1}
		\def\b{.3}
		\tikzstyle{nod1}= [circle, inner sep=0pt,fill=white,  minimum size=5pt, draw]
		\tikzstyle{nod}= [circle, inner sep=0pt, fill=black, minimum size=5pt, draw]
		% left triangle
		\coordinate (w00) at (0,0);
		\coordinate (w10) at (\l,0);
		\coordinate (w01) at (0,\l);
		\coordinate (w11) at (\l,\l);
		% middle square
		\coordinate (u00) at (\l+\d,0);
		\coordinate (u10) at (\l+\l+\d,0);
		\coordinate (u01) at (\l+\d,\l);
		\coordinate (u11) at (\l+\l+\d,\l);
		% right triangle
		\coordinate (v00) at (\l+\d+\d,0);
		\coordinate (v10) at (\l+\d+\d+\l,0);
		\coordinate (v01) at (\l+\d+\d,\l);
		\coordinate (v11) at (\l+\d+\d+\l,\l);

%		\draw[dashed] (\l+\l+\d+\le,-\le) --  (\l+\d-\le,\l+\le);
%		\draw[dashed] (\l+\d-\le,0-\le) --   (\l+\l+\d+\le,\l+\le);
		
		\draw[-] (w00) --  node [below]{ $\alpha$} (w10) -- (w11) -- (w01) -- node [left]{ $\beta$}  (w00) ;
		\draw[line width=1.5pt] (w00) --  (w11);
		\draw[line width=1.5pt] (v10) --  (v01);
		\node[nod1](w00) at (0,0)  [label=below: $x$] {};
		\node[nod1](w10) at (\l,0) [label=below: $y$] {};
		\node[nod1](w01) at (0,\l) [label=above: $c$ ] {};
		\node[nod1](w11) at (\l,\l) [label=above: $z$ ] {};

\node (q1) at (.5*\l,-2*\b) {$q(x,y,z;\alpha, \beta)=0$};
\node (q12) at (.5*\l,1.5*\b+\l) {$q(z,c,x;\alpha,\beta)=0$};

		\draw[-] (v10) --  node [below]{ $\alpha$} (v00) -- (v01) -- (v11) -- node [right]{ $\beta$} (v10) ;
		\node[nod1] (v10) at (\l+\d+\d+\l,0)[label=below:$y$ ] {};
		\node[nod1] (v00) at (\l+\d+\d,0) [label=below:$x$] {};
		\node[nod1] (v11) at (\l+\d+\d+\l,\l) [label=above:$z$] {};
		\node[nod1] (v01) at (\d+\d+\l,\l) [label=above: $c$ ] {};
		\node (q2) at (\l+\d+\d+.7*\l,-2*\b) {$p(y,x,c;\alpha, \beta)=0$};
	\node (q22) at (\l+\d+\d+.7*\l,\l+1.5*\b) {$p(c,z,y;\alpha, \beta)=0$};

		\node (1) at (0.7*\l+\l+.5*\d,.5*\l) {$ \Longleftrightarrow  $};

		\node (2) at (0.7*\l+\l+.5*\d,.2*\l) {$Q(x,y,c,z;\alpha,\beta)=0$};

%		\node (2) at (\l+.5*\d+\l+\d,.5*\l) {$\Leftrightarrow  $};
		
		\end{tikzpicture}
	\end{center}
	\caption{\label{fig:duality} Folding of $Q$ along the two diagonals: $p=0$ is the dual of $q=0$, and vice versa.  % The ${\cal D}4$ symmetry results in dual boundary equations characterised by $p$ and $q$, associated to the same involution $\sigma$
	}
\end{figure}

The duality property is illustrated in Figure~\ref{fig:duality}: if $Q$ is folded by either $p=0$ or $q=0$ then it is folded into two copies of $p$ through $q=0$ and vice versa.

It is important to note that the folding procedure explained here does not contain information about integrability of the boundary equation $q=0$. It is a first step towards selecting candidates for integrable $q$'s. Going back to the original construction of \cite{CCZ} involving the function $k$ as in \eqref{eq:k}, we see now that the latter is nothing but $-\frac{p_2}{p_1}$ obtained when eliminating $c$ using $p=0$. This function plays a crucial role in defining the boundary matrix $K$ appearing in the the boundary zero curvature equation (see section \ref{bd_ZC}). The notion of a pair of dual boundary equations $q=0$, $p=0$ puts on firm ground the idea of folding of $Q$ that was introduced in \cite{CCZ}. It is a valuable notion for (at least) two reasons: $1)$ it provides the precise link between the boundary consistency condition and the boundary zero curvature equation (see Proposition~\ref{th:bc1} below); $2)$ dual boundary equations provide good candidates for  integrable boundary equations. All the integrable boundary equations found in \cite{CCZ} fall into this category. In fact, many more integrable boundary equations for the ABS list can be obtained this way. This is left for future work.

\subsection{Dual boundary consistency and the boundary zero curvature condition}\label{bd_ZC}
A discrete boundary zero curvature condition, connecting a $3$D-consistent equation and its integrable boundary equations, was  formulated in \cite{CCZ}. In this section we explain the connection between boundary consistency and the discrete boundary zero curvature condition. Equipped with the notion of dual boundary equation, we now proceed to formulate a different boundary consistency condition in terms of a dual pair $(q,p)$.

\begin{definition}	
	\label{Def_dual_bd_consistent} Let $Q=0$ be an integrable quad equation, which admits a pair of dual boundary equations $q=0$, $p=0$. The ordered pair $(q,p)$ is  said to be {\bf dual boundary consistent with $Q=0$} if there is an involutive function $\sigma$ between the parameters, $\beta=\sigma(\alpha)$ and $\eta=\sigma(\lambda)$, such that the initial value problem on the $3$D-stencil in Figure~\ref{fig:dbccy}, where $p=0$ is imposed on the top and bottom triangles and $q=0$ is imposed on the vertical triangles, is well-posed, \ie the two ways of computing $e$ from initial values $x,y,u$ yield the same value.
\end{definition}
\begin{figure}[h]
% \hspace{5cm}
  \begin{center}
\begin{tikzpicture}[scale=.8]
      % \tikzstyle{nod1}= [circle, inner sep=0pt, minimum size=5pt, draw]
      \tikzstyle{nod1}= [circle, inner sep=0pt,fill=white,  minimum size=5pt, draw]
      \tikzstyle{nod}= [circle, inner sep=0pt, fill=black, minimum size=5pt, draw]
      \tikzstyle{vertex}=[circle,minimum size=20pt,inner sep=0pt]
      \tikzstyle{selected vertex} = [vertex, fill=red!24]
      \tikzstyle{selected edge} = [draw,line width=1.5pt,-]
      \tikzstyle{edge} = [draw, thin,-,black] %inner lines
      \tikzstyle{ddedge} = [draw, thick, dashed, -,black] %dotted lines
      \tikzstyle{dddedge} = [draw, thick, densely dotted, -,black] %dotted lines
      \tikzstyle{dedge} = [draw, dashed,-,black] %dashed lines
      \tikzstyle{eedge} = [draw, line width=1.5pt,-,black] %outer lines

      \pgfmathsetmacro \an {45}%projection angle
      \pgfmathsetmacro \lh {2.4}

      \pgfmathsetmacro \r {.34}%projection angle a
      \pgfmathsetmacro \L {3.3}% inner lenght
      \pgfmathsetmacro \Y {\L*\r}% inner lenght
      \pgfmathsetmacro \X {\Y*tan(\an)}% inner lenght

      \pgfmathsetmacro \ld {.9}
      \pgfmathsetmacro \R {.6}%projection angle a
      \pgfmathsetmacro \l {\L*\R}% inner lenght
      \pgfmathsetmacro \y {\Y*\R}
      \pgfmathsetmacro \x {\X*\R}
      % 1st floor v, v1, v4, v14
      \node[nod] (x) at (\L,0) [label=below:$x$] {};
      \node[nod] (y) at (0,0) [label=below:$y$] {};
      \node[nod1] (t) at (\X+\L,\Y) [label=right:$c$] {};
      \draw[edge] (y) -- node[below] {$\lambda$}  (x) -- node[right] {$\eta$}(t);
      % 3rd floor w, w1, w4, w14
      \node[nod1] (bx) at (\L,2*\lh) [label=above:$r$] {};
      \node[nod1] (by) at (0,0+2*\lh) [label=above:$s$] {};
      \node[nod1] (bt) at (\X+\L,\Y+2*\lh) [label=above:$e$] {};
      \draw[edge] (by) -- (bx) -- (bt);
    %   % 2nd floor u1,u2,u3,u4,u5,u6
      \node[nod] (u) at (\ld+\l,1.1*\lh) [label=above left:$u$] {};
      \node[nod1] (v) at (\ld,1.1*\lh) [label=left:$v$] {};
      \node[nod1] (T) at (\ld+\x+\l,\y+1.1*\lh) [label=right:$d$] {};
      \draw[edge] (v) -- (u) --  (T) ;
      \draw[edge] (by) -- (v) -- (y);
      \draw[edge] (bt) -- (T) -- (t);
      \draw[edge] (bx)  --node[above left] {$\beta$} (u) -- node[above left] {$\alpha$}(x);
      \draw[eedge] (x) -- (bx) ;
      \draw[eedge] (by)-- (bt) ;
      \draw[dashed, line width=1.5pt] (y)-- (t) ;
    \end{tikzpicture}
  \caption{Dual boundary consistency: the equation $q=0$ is imposed on the side vertical triangle and its dual $p=0$ is imposed on the top and bottom triangles. The two ways of computing $e$ from $x,y,u$ lead to the same value.}
  \label{fig:dbccy}
    \end{center}
\end{figure}
For boundary equations which admit a dual boundary equation the above consistency condition turns out to be equivalent to the original consistency condition.
\begin{proposition}\label{th:bc1}
A boundary equation $q = 0$, with dual $p = 0$, is boundary consistent with $Q = 0$ according to Definition \ref{Def_integ_bd}, if and only if the ordered pair $(q ,p)$ is dual boundary consistent with $Q = 0$ according to Definition \ref{Def_dual_bd_consistent}.
\end{proposition}
Note that the proposition does not imply that $p=0$ is boundary consistent if $q=0$ is.

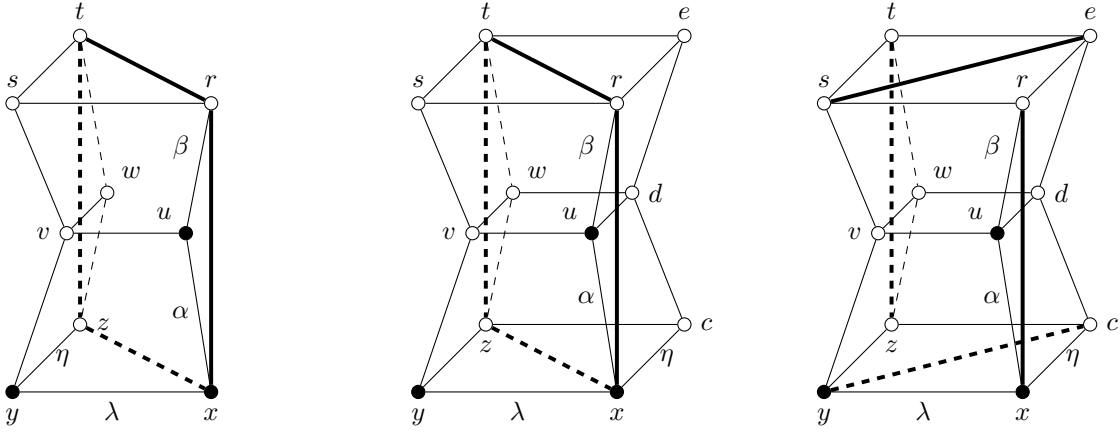
\begin{figure}[bt]
  \centering
% \hspace{2cm}
  \begin{subfigure}[b]{0.3\textwidth}
    \begin{tikzpicture}[scale=.8]
      \tikzstyle{nod1}= [circle, inner sep=0pt,fill=white,  minimum size=5pt, draw]
      \tikzstyle{nod}= [circle, inner sep=0pt, fill=black, minimum size=5pt, draw]
      \tikzstyle{vertex}=[circle,minimum size=20pt,inner sep=0pt]
      \tikzstyle{selected vertex} = [vertex, fill=red!24]
      \tikzstyle{selected edge} = [draw,line width=1.5pt,-]
      \tikzstyle{edge} = [draw, thin,-,black] %inner lines
      \tikzstyle{ddedge} = [draw, line width=1.5pt, dashed, -,black] %dotted lines
      \tikzstyle{dddedge} = [draw, thick, densely dotted, -,black] %dotted lines
      \tikzstyle{dedge} = [draw, dashed,-,black] %dashed lines
      \tikzstyle{eedge} = [draw, line width=1.5pt ,-,black] %outer lines
      \pgfmathsetmacro \an {45}%projection angle
      \pgfmathsetmacro \lh {2.4}
      \pgfmathsetmacro \r {.34}%projection angle a
      \pgfmathsetmacro \L {3.3}% inner lenght
      \pgfmathsetmacro \Y {\L*\r}% inner lenght
      \pgfmathsetmacro \X {\Y*tan(\an)}% inner lenght
      \pgfmathsetmacro \ld {.9}
      \pgfmathsetmacro \R {.6}%projection angle a
      \pgfmathsetmacro \l {\L*\R}% inner lenght
      \pgfmathsetmacro \y {\Y*\R}
      \pgfmathsetmacro \x {\X*\R}
      % 1st floor v, v1, v4, v14
      \node[nod] (x) at (\L,0) [label=below:$x$] {};
      \node[nod] (y) at (0,0) [label=below:$y$] {};
      \node[nod1] (z) at (\X,\Y) [label=right:$z$] {};
     % \node[nod1] (t) at (\X+\L,\Y) [label=right:$t$] {};
      \draw[edge] (x) -- node[below] {$\lambda$}  (y) -- node[right] {$\eta$}  (z);% -- (t) --node[right] {$\eta$}(x) ;
      % 3rd floor w, w1, w4, w14
      \node[nod1] (bx) at (\L,2*\lh) [label=above:$r$] {};
      \node[nod1] (by) at (0,0+2*\lh) [label=above:$s$] {};
      \node[nod1] (bz) at (\X,\Y+2*\lh) [label=above:$t$] {};
     % \node[nod1] (bt) at (\X+\L,\Y+2*\lh) [label=above:$e$] {};
      \draw[edge] (bx) -- (by) -- (bz);% -- (bt) --(bx) ;
    %   % 2nd floor u1,u2,u3,u4,u5,u6
      \node[nod] (u) at (\ld+\l,1.1*\lh) [label=above left:$u$] {};
      \node[nod1] (v) at (\ld,1.1*\lh) [label=left:$v$] {};
      \node[nod1] (w) at (\ld+\x,\y+1.1*\lh) [label=above right:$w$] {};
     % \node[nod1] (T) at (\ld+\x+\l,\y+1.1*\lh) [label=right:$\tau$] {};
      \draw[edge] (u) -- (v) -- (w);% -- (T) --(u) ;
      \draw[edge] (by) -- (v) -- (y);
     % \draw[edge] (bt) -- (T) -- (t);
      \draw[edge] (bx) -- node[above left] {$\beta$} (u) -- node[left] {$\alpha$}  (x);
      \draw[eedge] (x) -- (bx) -- (bz);
      \draw[dedge] (bz)-- (w) -- (z);
      \draw[ddedge] (bz)-- (z) -- (x);
      % \draw[densely dotted, thick] (by)-- (bt) ;
      % \draw[densely dotted, thick] (y)-- (t) ;
    \end{tikzpicture}
  \end{subfigure}
  %\hspace{.5cm}
  \begin{subfigure}[b]{0.3\textwidth}
    \begin{tikzpicture}[scale=.8]
      \tikzstyle{nod1}= [circle, inner sep=0pt,fill=white,  minimum size=5pt, draw]
      \tikzstyle{nod}= [circle, inner sep=0pt, fill=black, minimum size=5pt, draw]
      \tikzstyle{vertex}=[circle,minimum size=20pt,inner sep=0pt]
      \tikzstyle{selected vertex} = [vertex, fill=red!24]
      \tikzstyle{selected edge} = [draw,line width=1.5pt,-]
      \tikzstyle{edge} = [draw, thin,-,black] %inner lines
      \tikzstyle{ddedge} = [draw, line width=1.5pt, dashed, -,black] %dotted lines
      \tikzstyle{dddedge} = [draw, thick, densely dotted, -,black] %dotted lines
      \tikzstyle{dedge} = [draw, dashed,-,black] %dashed lines
      \tikzstyle{eedge} = [draw, line width=1.5pt,-,black] %outer lines
      \pgfmathsetmacro \an {45}%projection angle
      \pgfmathsetmacro \lh {2.4}
      \pgfmathsetmacro \r {.34}%projection angle a
      \pgfmathsetmacro \L {3.3}% inner lenght
      \pgfmathsetmacro \Y {\L*\r}% inner lenght
      \pgfmathsetmacro \X {\Y*tan(\an)}% inner lenght
      \pgfmathsetmacro \ld {.9}
      \pgfmathsetmacro \R {.6}%projection angle a
      \pgfmathsetmacro \l {\L*\R}% inner lenght
      \pgfmathsetmacro \y {\Y*\R}
      \pgfmathsetmacro \x {\X*\R}
      % 1st floor v, v1, v4, v14
      \node[nod] (x) at (\L,0) [label=below:$x$] {};
      \node[nod] (y) at (0,0) [label=below:$y$] {};
      \node[nod1] (z) at (\X,\Y) [label=below:$z$] {};
      \node[nod1] (t) at (\X+\L,\Y) [label=right:$c$] {};
      \draw[edge] (x) -- node[below] {$\lambda$}  (y) -- (z) -- (t) --node[right] {$\eta$}(x) ;
      % 3rd floor w, w1, w4, w14
      \node[nod1] (bx) at (\L,2*\lh) [label=above:$r$] {};
      \node[nod1] (by) at (0,0+2*\lh) [label=above:$s$] {};
      \node[nod1] (bz) at (\X,\Y+2*\lh) [label=above:$t$] {};
      \node[nod1] (bt) at (\X+\L,\Y+2*\lh) [label=above:$e$] {};
      \draw[edge] (bx) -- (by) -- (bz) -- (bt) --(bx) ;
    %   % 2nd floor u1,u2,u3,u4,u5,u6
      \node[nod] (u) at (\ld+\l,1.1*\lh) [label=above left:$u$] {};
      \node[nod1] (v) at (\ld,1.1*\lh) [label=left:$v$] {};
      \node[nod1] (w) at (\ld+\x,\y+1.1*\lh) [label=above right:$w$] {};
      \node[nod1] (T) at (\ld+\x+\l,\y+1.1*\lh) [label=right:$d$] {};
      \draw[edge] (u) -- (v) -- (w) -- (T) --(u) ;
      \draw[edge] (by) -- (v) -- (y);
      \draw[edge] (bt) -- (T) -- (t);
      \draw[edge] (bx) -- node[above left] {$\beta$} (u) -- node[above left] {$\alpha$}  (x);
      \draw[eedge] (x) -- (bx) -- (bz);
      \draw[dedge] (bz)-- (w) -- (z);
      \draw[ddedge] (bz)-- (z) -- (x);
      % \draw[densely dotted, thick] (by)-- (bt) ;
      % \draw[densely dotted, thick] (y)-- (t) ;
    \end{tikzpicture}
  \end{subfigure}
    \begin{subfigure}[b]{0.3\textwidth}
    \begin{tikzpicture}[scale=.8]
      \tikzstyle{nod1}= [circle, inner sep=0pt,fill=white,  minimum size=5pt, draw]
      \tikzstyle{nod}= [circle, inner sep=0pt, fill=black, minimum size=5pt, draw]
      \tikzstyle{vertex}=[circle,minimum size=20pt,inner sep=0pt]
      \tikzstyle{selected vertex} = [vertex, fill=red!24]
      \tikzstyle{selected edge} = [draw,line width=1.5pt,-]
      \tikzstyle{edge} = [draw, thin,-,black] %inner lines
      \tikzstyle{ddedge} = [draw, line width=1.5pt, dashed, -,black] %dotted lines
      \tikzstyle{dddedge} = [draw, thick, densely dotted, -,black] %dotted lines
      \tikzstyle{dedge} = [draw, dashed,-,black] %dashed lines
      \tikzstyle{eedge} = [draw, line width=1.5pt,-,black] %outer lines
      \pgfmathsetmacro \an {45}%projection angle
      \pgfmathsetmacro \lh {2.4}
      \pgfmathsetmacro \r {.34}%projection angle a
      \pgfmathsetmacro \L {3.3}% inner lenght
      \pgfmathsetmacro \Y {\L*\r}% inner lenght
      \pgfmathsetmacro \X {\Y*tan(\an)}% inner lenght
      \pgfmathsetmacro \ld {.9}
      \pgfmathsetmacro \R {.6}%projection angle a
      \pgfmathsetmacro \l {\L*\R}% inner lenght
      \pgfmathsetmacro \y {\Y*\R}
      \pgfmathsetmacro \x {\X*\R}
      % 1st floor v, v1, v4, v14
      \node[nod] (x) at (\L,0) [label=below:$x$] {};
      \node[nod] (y) at (0,0) [label=below:$y$] {};
      \node[nod1] (z) at (\X,\Y) [label=below:$z$] {};
      \node[nod1] (t) at (\X+\L,\Y) [label=right:$c$] {};
      \draw[edge] (x) -- node[below] {$\lambda$}  (y) -- (z) -- (t) --node[right] {$\eta$}(x) ;
      % 3rd floor w, w1, w4, w14
      \node[nod1] (bx) at (\L,2*\lh) [label=above:$r$] {};
      \node[nod1] (by) at (0,0+2*\lh) [label=above:$s$] {};
      \node[nod1] (bz) at (\X,\Y+2*\lh) [label=above:$t$] {};
      \node[nod1] (bt) at (\X+\L,\Y+2*\lh) [label=above:$e$] {};
      \draw[edge] (bx) -- (by) -- (bz) -- (bt) --(bx) ;
    %   % 2nd floor u1,u2,u3,u4,u5,u6
      \node[nod] (u) at (\ld+\l,1.1*\lh) [label=above left:$u$] {};
      \node[nod1] (v) at (\ld,1.1*\lh) [label=left:$v$] {};
      \node[nod1] (w) at (\ld+\x,\y+1.1*\lh) [label=above right:$w$] {};
      \node[nod1] (T) at (\ld+\x+\l,\y+1.1*\lh) [label=right:$d$] {};
      \draw[edge] (u) -- (v) -- (w) -- (T) --(u) ;
      \draw[edge] (by) -- (v) -- (y);
      \draw[edge] (bt) -- (T) -- (t);
      \draw[edge] (bx) -- node[above left] {$\beta$} (u) -- node[above left] {$\alpha$}  (x);
      \draw[eedge] (x) -- (bx);
      \draw[dedge] (bz)-- (w) -- (z);
      \draw[ddedge] (bz)-- (z);
       \draw[eedge] (by)-- (bt) ;
       \draw[ddedge] (y)-- (t) ;
    \end{tikzpicture}
  \end{subfigure}
  \caption{The equivalence between boundary consistency and dual boundary consistency through duality property.}
  \label{fig:bzcc1}
\end{figure}

\medskip
\noindent
\prf
For convenience, when $Q=0$ is used to express one variable in terms of the other three (and the parameters), we will write for short $(x,y,u)\underset{Q}{\longrightarrow} v$. Similarly, with $q=0$ we write for instance $(x,u)\underset{q}{\longrightarrow} r$. Recall that $\beta=\sigma(\alpha)$ and $\eta=\sigma(\lambda)$.
\begin{itemize}
\item[$\Rightarrow$]
Consider the figure on the left in Figure~\ref{fig:bzcc1} which represents the boundary consistency where  $q=0$ is imposed on the four boundary triangles and $Q=0$ is imposed on the four quadrilaterals. The values of the vertices $z,v,w,r,s,t$ are consistently defined. We embed it into the middle figure by adding the vertices $c,d,e$ which are defined as follows: $(x,y,z)\underset{Q}{\longrightarrow} c$, $(u,v,w)\underset{Q}{\longrightarrow} d$ and $(r,s,t)\underset{Q}{\longrightarrow} e$. The equation $Q=0$ is imposed on the added quadrilaterals $(xcdu)$, $(czwd)$, $(uder)$ and $(dwte)$. The $3$D consistency of $Q=0$ ensures that $d$ and $e$ are defined uniquely and consistently. Finally, we move to the figure on the right by noting that on the bottom quadrilateral, we now have $q(x,y,z;\lambda,\eta)=0$ and $Q(x,y,c,z;\lambda,\eta)=0$ so that by the duality property, we have $p(y,x,c;\lambda,\eta)=0$. This is indicated by the change of the dashed diagonal line from $(xz)$ to $(yc)$. Similarly, on the top quadrilateral, we have $p(s,r,e;\lambda,\eta)=0$.

From the point of view of the initial value problem, we can determine $e$ consistently from $y,x,u$ as follows:

$$(x,y,u)\underset{Q}{\longrightarrow} v\,,~~(x,u)\underset{q}{\longrightarrow} r\,,~~(u,v,r)\underset{Q}{\longrightarrow} s\,,~~(s,r)\underset{p}{\longrightarrow} e\,,$$
or
$$(x,y)\underset{p}{\longrightarrow} c\,,~~(c,x,u)\underset{Q}{\longrightarrow} d\,,~~(x,u)\underset{q}{\longrightarrow} r\,,~~(d,u,r)\underset{Q}{\longrightarrow} e\,.$$

It remains to delete the vertices $z,w,t$ to obtain precisely the dual boundary consistency of $(q,p)$ with $Q$ illustrated in Figure~\ref{fig:dbccy}.

\item[$\Leftarrow$] Starting from the dual boundary consistency diagram~\ref{fig:dbccy} and  adding the vertices $z$, $w$ and $t$ by using $Q$ as before, we obtain the figure on the right-hand in Figure~\ref{fig:bzcc1} where all the vertices are consistently defined from $x,y,u$. The factorisation of $Q$ implies we have $p(y,z,c;\eta,\lambda)=p(s,t,e;\eta,\lambda)=0$ and the dual boundary consistency of $(q,p)$ with $Q=0$ implies that $q(z,w,t;\alpha,\beta)=0$. We can switch the diagonals on the top and bottom faces due to the duality property between $p$ and $q$. Lastly, we delete the vertices $c,d,e$ to obtain the boundary consistency condition between $q$ and $Q$, \ie the figure on the left-hand side, as desired. \finprf
\end{itemize}

The significance of Proposition~\ref{th:bc1} is that, through the dual object $p$, we can relate the boundary consistency for $q$ and $Q$ with a discrete boundary zero curvature condition involving a boundary matrix $K$, in the same way as one can relate the $3$D consistency of $Q$ with the zero curvature condition \eqref{eq:3mllm1}. We now explain how this works.

Since integrable boundary equations only depend on one parameter, as the second parameter $\eta =\sigma(\lambda)$ is related to the first by the involution $\sigma$, in the sequel we write $q=q(x,y,z;\lambda)$, and similarly for $p$. As explained in Section \ref{3D_cons}, we can associate a Lax matrix $L$ to $Q$. Similarly, we can associate a boundary matrix $K$ to $p(y,x,c;\lambda) =0$, by expressing $c$ as a M\"obius transformation acting on $y$:
\begin{equation}
  \label{eq:4}
c= \frac{k_1(x;\lambda) y+ k_2(x;\lambda)}{k_3(x;\lambda) y+ k_4(x;\lambda)}= K \,[y]  \,, \quad K=K(x;\lambda)=  \kappa K_c
\end{equation}
where $\kappa$ is a scalar function and $K_c=\bma k_1 &  k_2 \\ k_3 & k_4\ema$ is the `core' of the {\bf  boundary matrix} $K$. The $\ZZ_2$ symmetry of $p$ implies that
$K_c(x;\lambda)K_c(x;\sigma(\lambda))$ is proportional to the identity matrix.

Using Figure~\ref{fig:dbccy},
by composition of M\"obius transformations,  we get the following {\em projective} discrete boundary zero curvature condition showing the two ways of expressing $e$ from $y$
\begin{equation}
\label{eq:3mllm}
L_c(r, u,\sigma(\alpha), \sigma(\lambda) )\, L_c(u,x,\alpha, \sigma(\lambda))\,
K_c(x,\lambda)\,[y] =
K_c(r;\lambda)\,
L_c(r, u;\sigma(\alpha), \lambda)\, L_c(u,x;\alpha,\lambda)\,[y],
\end{equation}
which holds when $q(x,u,r;\alpha)=0$, with $\beta = \sigma(\alpha)$, cf. property (iv) for $q$. As was the case for the bulk equation $Q=0$, we want to make the boundary zero curvature condition into a true matrix equation which should hold on $q=0$. Letting $u=\widetilde{x}$ and $r=\widehat{\widetilde{x}}$, and
\begin{equation}
L=L(\widetilde{x}, x ; \alpha, \lambda),\quad
M=L(\widehat{x}, x ; \sigma(\alpha), \lambda),
\end{equation}
then equation (\ref{eq:3mllm}) can be written as
$ \widetilde{\overline{M}}\, \overline{L}\, K\, [y] =
\widehat{\widetilde{K}}\, \widetilde{M}\, L\, [y] $,
whose matrix version reads
\begin{equation} \label{mbzcc}
\widetilde{\overline{M}}\, \overline{L}\, K=
\widehat{\widetilde{K}}\, \widetilde{M}\, L\,.
\end{equation}
Taking determinants, using normalisation \eqref{eq:invL} and setting for convenience $\rho=\sqrt{\det{K}}=\kappa\sqrt{\det{K_c}}$, we obtain the following condition relating the normalisation of $L$ and that of $K$,
\begin{equation}
\left(\frac{\widehat{\widetilde{\rho}}}{\rho}\right)^2=
\frac{\ell(\sigma(\alpha),\sigma(\lambda))\, \ell(\alpha,\sigma(\lambda))}{\ell(\sigma(\alpha),\lambda) \,\ell(\alpha,\lambda)}\,.
\end{equation}
Note that the left-hand side is independent of $\alpha$ so can be evaluated at $\alpha=\alpha_0$ where $\alpha_0$ is a fixed point of $\sigma$. This implies that
\begin{equation}
\frac{\widehat{\widetilde{\rho}}}{\rho}=\epsilon
\frac{ \ell(\alpha_0,\sigma(\lambda))}{\ell(\alpha_0,\lambda)}\,,~~ \epsilon^2=1\,.
\end{equation}
We can now re-scale $L$ by a function of $\lambda$ as
\begin{equation}
\label{rescaleL1}
L(u,x;a,\lambda)\to\frac{L(u,x;a,\lambda)}{\sqrt{\ell(\alpha_0,\lambda)}}\,,~~ 
\quad \ell(\alpha,\lambda)\to \frac{\ell(\alpha,\lambda)}{\ell(\alpha_0,\lambda)},
\end{equation}
to obtain
\begin{equation}
\label{rho}
\widehat{\widetilde{\rho}}=\epsilon \rho,\qquad \epsilon^2=1\,.
\end{equation}
This suggests that $K$ should be normalised such that $\rho$ does not depend on the field $x$ as otherwise \eqref{rho} could lead to a relation between $x$ and $r$ which is incompatible with $q(x,u,r;\alpha)=0$. In practice, we normalise the matrix $K$ such that 
\begin{equation}
\label{KK}
K(x;\lambda)\,K(x;\sigma(\lambda))=\text{id}\,.
\end{equation} 
The reason for not scaling $L$ to have determinant $1$ is to avoid unnecessary square roots which are hard to deal with in a computer algebra environment. Allowing $\epsilon$ to be $\pm 1$ is necessary  for (\ref{mbzcc}) to be equivalent to the boundary equation in general. Indeed, in some examples (see below), it can happen that $\epsilon= -1$ is required. 
With this in mind, and summarising our account, we will take
\begin{equation}
\label{KLLfinal}
L(r, u;\sigma(\alpha), \sigma(\lambda) )\, L(u,x;\alpha, \sigma(\lambda))\,
K(x;\lambda) =\epsilon\,
K(r;\lambda)\,
L(r, u;\sigma(\alpha), \lambda)\, L(u,x;\alpha,\lambda)\,,\quad \epsilon=\pm 1\,,
\end{equation}
as the boundary zero curvature representation of the integrable boundary equation $q(x,u,r;\alpha)=0$ in the rest of this paper, with the understanding that $L$ is normalised by applying the rescaling \eqref{rescaleL1} to \eqref{eq:invL} and $K$ is normalised as in \eqref{KK}, with a determinant independent of the field.

In general, both $L$ and $K$ may involve square roots of the  parameters and/or fields in their expressions, which makes it difficult to extract invariants. However, in certain cases the freedom we have exploited here makes it possible to deal with this. This will be clear in the examples below.

%This concludes our account on the relation between boundary consistency and boundary zero curvature equation. These notions were originally presented in \cite{CCZ} but without the clear connection obtained here which relies on the important Theorem \ref{th:bc1} and Figure~\ref{fig:dbccy}. The boundary zero curvature representation of an integrable boundary equation is of central importance for our new procedure of open boundary reductions explained in Section \ref{open_red}.

\subsection{Examples of integrable boundary equations and boundary matrices}\label{sec:ex}

We provide integrable boundary equations and their boundary matrices for the H1 and Q1($\delta=0$) equations from the ABS list. These examples will be used in Section~$4$ in the  construction of integrable mappings by the open reduction method. Note that for all integrable boundary equations obtained in \cite{CCZ}, the involution $\sigma$ is either in an ``additive'' form:
\begin{equation}\label{eq:addp}
  \sigma (\alpha) = -\alpha +2\mu\,,
\end{equation}
or in a  ``multiplicative'' form:
\begin{equation}\label{eq:mulp}
  \sigma (\alpha) = \frac{\mu^2}{\alpha}\,,
\end{equation}
where $\mu$ is a free parameter in both cases ($\mu \neq 0$ in the multiplicative case). In each case, $\alpha_0=\mu$ is a fixed point of $\sigma$ which we use to implement \eqref{rescaleL1}.

In order to illustrate the procedure of normalization for the boundary zero curvature equation, for each example we give $L$ corresponding to $Q=0$ with the normalization obtained in \cite{BHQK} and the corresponding function $\ell(\alpha,\lambda)$. For all examples considered below, the boundary matrix $K$ satisfies that  $\det K$ is a constant, then  it
is understood that,
given $\sigma$ as in \eqref{eq:addp}, or \eqref{eq:mulp}, one should then perform \eqref{rescaleL1}.

\subsubsection{H1, case of additive $\sigma$}
The equation reads
\begin{equation}
\label{eq:H1}
(u-\widehat{\widetilde{u}})(\widetilde{u}-\widehat{u})+\beta-\alpha=0\,,
\end{equation}
and has Lax matrix
\begin{equation}
L(\widetilde{u},u;\alpha,\lambda)=
  \begin{pmatrix}
    u & \alpha-\lambda -u \widetilde{ u } \\ 1 & -\widetilde{ u }
  \end{pmatrix}\,,\quad \ell (\alpha,\lambda) = \lambda-\alpha\,,
\end{equation}
Table~\ref{Tabh1} gives the required elements to obtain the boundary zero curvature representation of $q=0$ for each example. Here, $\sigma(\alpha) =-\alpha+2 \mu$, with $\mu$ being a free parameter appearing also in one of the integrable boundary equations. % {\color{blue} As to the  constraint \eqref{req}, one can check that $\det K$ is a constant for both equations, and the left-hand side of \eqref{req} is $1$, \ie
% \begin{equation}
%   \frac{\ell(\sigma(\alpha),\lambda) \,\ell(\alpha,\lambda)}{\ell(\sigma(\alpha),\sigma(\lambda))\, \ell(\alpha,\sigma(\lambda))} = 1\,,\quad \sigma(\alpha) = -\alpha+2\mu\,.
% \end{equation}
% }

\begin{table}[h!]
  \begin{center}
    \begin{tabular}{ c  | c |  c | c c c}
    {boundary equation}& dual boundary equation &  \multicolumn{2}{c } { boundary matrix} \\
      \hline
                         $q(x,y,z;\alpha)=0$ &$p(y,x,c,\lambda)=0$&  $ \epsilon$  & $K(x;\lambda)$\\ %
      \hline                                     $y(z-x)+\alpha-\mu=0$  & $y+c=0$ &  $1$ &  $\bma -1 & 0  \\ 0 & 1\ema$ \\
                                   $x+z=0$  &$x(y-c)+\mu-a=0$ &  $-1$ & $\bma 1 & \frac{\mu-\lambda}{x}  \\ 0 & 1\ema$\\
                  \end{tabular}
    \caption{Integrable boundary equations for H1.}
    \label{Tabh1}
  \end{center}
\end{table}

\subsubsection{Q1($\delta=0$), case of additive $\sigma$}
The equation reads
\begin{equation}
  \label{eq:q101}
  \alpha (u - \widehat{u}) (\widetilde{u} - \widehat{\widetilde{u}}) - \beta (u - \widetilde{u}) (\widehat{u} - \widehat{\widetilde{u}}) =0\,,
\end{equation}
with Lax matrix
\begin{equation}
L(\widetilde{u},u;\alpha,\lambda) = \frac{1}{\widetilde{u}-u}\bma \lambda(\widetilde{u}-u) -\alpha \widetilde{u}& \alpha \widetilde{u} u \\   -\alpha & \lambda(\widetilde{u}-u)+ \alpha {{u}}\ema\,,\quad \ell(\alpha,\lambda) = \lambda(\lambda-\alpha)\,.
\end{equation}
 Table~\ref{Tabq1} gives the required elements to obtain the boundary zero curvature representation of $q=0$ for each example. Here, $\sigma(\alpha)=-\alpha+2\mu$.
\begin{table}[h!]
	\begin{center}
		\begin{tabular}{   c | c |  c | c c c}
			{boundary equation}& dual boundary equation &  \multicolumn{2}{c }{boundary matrix} \\
			\hline
		 $q(x,y,z;\alpha)=0$ &$p(y,x,c,\lambda)=0$&  $ \epsilon$ &  $K(x;\lambda)$\\ %
			\hline
		 $\alpha(x z-y^2 ) - \mu(x-y) (y+z)  = 0$  & $y+c=0$ &  $1$ &  $\bma -1 & 0  \\ 0 & 1\ema$ \\
			 $x+z=0$  &$ a(x^2-cy)+\mu(c+x)(y-x)=0$ &  $-1$ &  $\frac{1}{\lambda(2\mu-\lambda)}\bma \mu  & (\lambda-\mu)x  \\ \frac{(\lambda-\mu)}{x} & \mu \ema$ \\
		\end{tabular}
		\caption{Examples of integrable boundary equations for Q1($\delta=0$) under additive $\sigma$. }
		\label{Tabq1}
	\end{center}
\end{table}

\subsubsection{Q1($\delta=0$), case of multiplicative $\sigma$}
For convenience, let us consider another form of Q1($\delta=0$)
\begin{equation}
  \label{eq:q102}
  \frac{1}{\alpha^2} (u - \widehat{u}) (\widetilde{u} - \widehat{\widetilde{u}}) - \frac{1}{\beta^2} (u - \widetilde{u}) (\widehat{u} - \widehat{\widetilde{u}}) =0\,,
\end{equation}
which has Lax matrix
\begin{equation}
\label{LQ1}
L(\widetilde{u},u;\alpha,\lambda) = \frac{1}{\alpha^2(\widetilde{u}-u)}\bma \alpha^2(\widetilde{u}-u) -\lambda^2 \widetilde{u}& \lambda^2 \widetilde{u} u \\   -\lambda^2 & \alpha^2(\widetilde{u}-u)+ \lambda^2 {{u}}\ema\,,\quad \ell(\alpha,\lambda) = 1-\frac{\lambda^2}{\alpha^2}\,.
\end{equation}
Here, $\sigma(\alpha) =\mu^2/\alpha$.
\begin{table}[h!]
	\begin{center}
		\begin{tabular}{   c | c |  c | c c c}
			{boundary equation}& dual boundary equation &  \multicolumn{2}{c }{boundary matrix} \\
			\hline
			$q(x,y,z;\alpha)=0$ &$p(y,x,c,\lambda)=0$&  $ \epsilon$ &  $K(x;\lambda)$\\ %
			\hline
			$\alpha^2(x-y)+\mu^2(y-z)=0$  & $a^2(y-x)-\mu^2(x-c)=0$ &  $1$ &  $\bma -\frac{\lambda}{\mu} & \frac{(\lambda^2+\mu^2)x}{\lambda\mu}   \\ 0 & \frac{\mu}{\lambda}\ema$ \\
			$\alpha^2(x-y)-\mu^2(y-z)=0$  &$ a^2(y-x)+\mu^2(x-c)=0$ &  $-1$ &  $\bma \frac{\lambda}{\mu} & \frac{(\mu^2-\lambda^2)x}{\lambda\mu}   \\ 0 & \frac{\mu}{\lambda}\ema$ \\
			\hline
			$\alpha^2(x-y)z+\mu^2(y-z)x=0$  & $a^2(y-x)c-\mu^2(x-c)y=0$ &  $1$ &  $\bma -\frac{\mu}{\lambda} & 0   \\ -\frac{\lambda^2+\mu^2}{\lambda\mu x} & \frac{\lambda}{\mu} \ema$ \\
			$\alpha^2(x-y)z-\mu^2(y-z)x=0$  & $a^2(y-x)c+\mu^2(x-c)y=0$ &  $-1$ &  $\bma \frac{\mu}{\lambda} & 0   \\ \frac{\mu^2-\lambda^2}{\lambda\mu x} & \frac{\lambda}{\mu} \ema$ \\
		\end{tabular}
		    \caption{Examples of integrable boundary equations for  Q1($\delta=0$) as in \eqref{eq:q102} with a multiplicative $\sigma$.}
		\label{Tabq2}
	\end{center}
\end{table}

\section{Integrable mappings from open boundary reductions}\label{open_red}
We introduce the idea of open boundary reductions of quad equations on quad-graphs with two parallel boundaries as a new means to construct integrable mappings. The key ingredients are the well-posedness of the initial-boundary data as well as the boundary zero curvature conditions.  A generating function for the invariants of these mappings will be obtained.

\subsection{Open boundary reductions on the $\ZZ^2$-lattice}

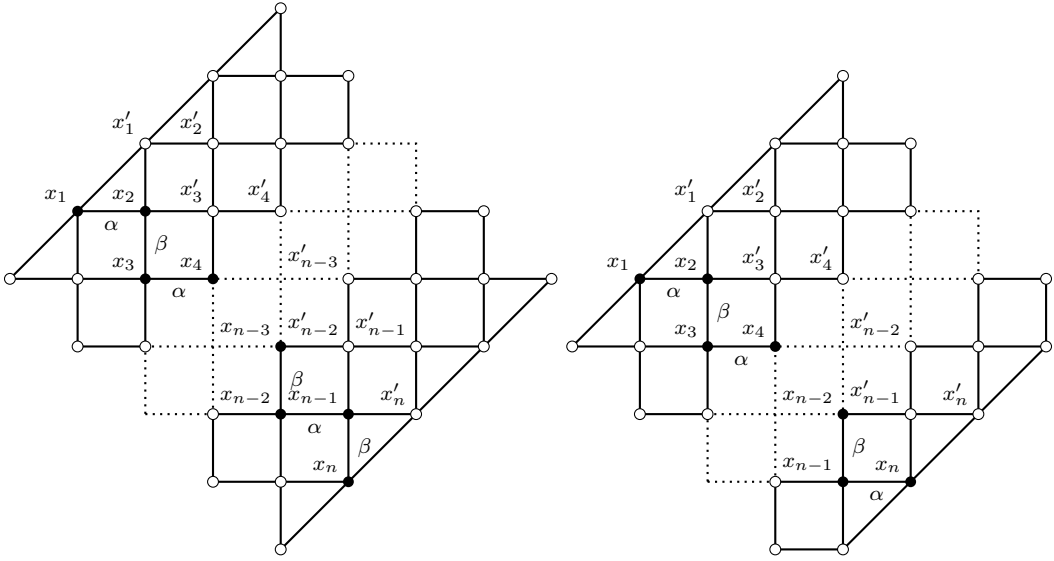
\begin{figure}[h!]
\begin{center}
\begin{tikzpicture}[scale=0.9]
\draw[thick]
    (0,0)--(4,4)--(4,1)--(1,1)--(1,-1)--(2,-1)--(2,2)--(5,2)--(5,3)
    --(5,3)--(3,3)--(3,0)--(0,0)
    (4,-4)--(8,0)--(5,0)--(5,-3)--(3,-3)--(3,-2)--(6,-2)--(6,1)--(7,1)
    --(7,-1)--(4,-1)--(4,-4);
\draw[dotted,thick]
    (2,-2)--(3,-2)--(3,0)--(5,0)--(5,2)--(6,2)--(6,1)--(4,1)--(4,-1)--(2,-1)--(2,-2);
\fill
    \foreach \x in {(1,1),(2,1),(2,0),(3,0),(4,-1),(4,-2),(5,-2),(5,-3)}
            {\x circle (0.08cm)};
\fill[color=white]
    \foreach \x in {(0,0),(1,0),(1,-1),(2,-1),(3,-2),(3,-3),(4,-3),(4,-4),(4,4),(3,3),(4,3),
    (5,3),(2,2),(3,2),(4,2),(5,2),(3,1),(4,1),(6,-2),(5,-1),(6,-1),(7,-1),(5,0),
    (6,0),(7,0),(8,0),(6,1),(7,1)}
            {\x circle (0.08cm)};
\draw
    \foreach \x in {(0,0),(1,0),(1,-1),(2,-1),(3,-2),(3,-3),(4,-3),(4,-4),(4,4),(3,3),(4,3),
    (5,3),(2,2),(3,2),(4,2),(5,2),(3,1),(4,1),(6,-2),(5,-1),(6,-1),(7,-1),(5,0),
    (6,0),(7,0),(8,0),(6,1),(7,1)}
            {\x circle (0.08cm)};
\draw   (1,1) node[above left] {\footnotesize $x_1$}
        (2,1) node[above left] {\footnotesize$x_2$}
        (2,0) node[above left] {\footnotesize$x_3$}
        (3,0) node[above left] {\footnotesize$x_4$}
        (4,-1) node[above left] {\footnotesize$x_{n-3}$}
        (4,-2) node[above left] {\footnotesize$x_{n-2}$}
        (5,-2) node[above left] {\footnotesize$x_{n-1}$}
        (5,-3) node[above left] {\footnotesize$x_{n}$};
\draw   (2,2) node[above left] {\footnotesize$x_1^\prime$}
        (3,2) node[above left] {\footnotesize$x_2^\prime$}
        (3,1) node[above left] {\footnotesize$x_3^\prime$}
        (4,1) node[above left] {\footnotesize$x_4^\prime$}
        (5,0) node[above left] {\footnotesize$x_{n-3}^\prime$}
        (5,-1) node[above left] {\footnotesize$x_{n-2}^\prime$}
        (6,-1) node[above left] {\footnotesize$x_{n-1}^\prime$}
        (6,-2) node[above left] {\footnotesize$x_{n}^\prime$};
\draw   (1.5,1) node[below] {\footnotesize$\alpha$}
        (2,.5) node[right] {\footnotesize$\beta$}
        (2.5,0) node[below] {\footnotesize$\alpha$}
        (4,-1.5) node[right] {\footnotesize$\beta$}
        (4.5,-2) node[below] {\footnotesize$\alpha$}
        (5,-2.5) node[right] {\footnotesize$\beta$};

      \end{tikzpicture}
\begin{tikzpicture}[scale=0.9]
\draw[thick]
    (0,0)--(4,4)--(4,1)--(1,1)--(1,-1)--(2,-1)--(2,2)--(5,2)--(5,3)
    --(5,3)--(3,3)--(3,0)--(0,0)
    (4,-3)--(7,0)--(5,0)--(5,-2)--(3,-2)--(3,-3)--(4,-3)--(4,-1)--(6,-1)
    --(6,1)--(7,1)--(7,0);
\draw[dotted,thick]
    (2,-2)--(3,-2)--(3,0)--(5,0)--(5,2)--(6,2)--(6,1)--(4,1)--(4,-1)--(2,-1)--(2,-2);
\fill
    \foreach \x in {(1,1),(2,1),(2,0),(3,0),(4,-1),(4,-2),(5,-2)}
            {\x circle (0.08cm)};
\fill[color=white]
    \foreach \x in {(0,0),(1,0),(1,-1),(2,-1),(3,-2),(3,-3),(4,-3),(4,4),(3,3),(4,3),
    (5,3),(2,2),(3,2),(4,2),(5,2),(3,1),(4,1),(5,-1),(6,-1),(5,0),
    (6,0),(7,0),(6,1),(7,1)}
            {\x circle (0.08cm)};
\draw
    \foreach \x in {(0,0),(1,0),(1,-1),(2,-1),(3,-2),(3,-3),(4,-3),(4,4),(3,3),(4,3),
    (5,3),(2,2),(3,2),(4,2),(5,2),(3,1),(4,1),(5,-1),(6,-1),(5,0),
    (6,0),(7,0),(6,1),(7,1)}
            {\x circle (0.08cm)};
\draw   (1,1) node[above left] {\footnotesize$x_1$}
        (2,1) node[above left] {\footnotesize$x_2$}
        (2,0) node[above left] {\footnotesize$x_3$}
        (3,0) node[above left] {\footnotesize$x_4$}
        (4,-1) node[above left] {\footnotesize$x_{n-2}$}
        (4,-2) node[above left] {\footnotesize$x_{n-1}$}
        (5,-2) node[above left] {\footnotesize$x_{n}$};
\draw   (2,2) node[above left] {\footnotesize$x_1^\prime$}
        (3,2) node[above left] {\footnotesize$x_2^\prime$}
        (3,1) node[above left] {\footnotesize$x_3^\prime$}
        (4,1) node[above left] {\footnotesize$x_4^\prime$}
        (5,0) node[above left] {\footnotesize$x_{n-2}^\prime$}
        (5,-1) node[above left] {\footnotesize$x_{n-1}^\prime$}
        (6,-1) node[above left] {\footnotesize$x_{n}^\prime$};
\draw   (1.5,1) node[below] {\footnotesize$\alpha$}
        (2,.5) node[right] {\footnotesize$\beta$}
        (2.5,0) node[below] {\footnotesize$\alpha$}
        (4,-1.5) node[right] {\footnotesize$\beta$}
        (4.5,-2) node[below] {\footnotesize$\alpha$};
\end{tikzpicture}
\end{center}
\caption{\label{fig1} A well-posed initial-boundary-value problem on $\ZZ^2$-lattice  on a strip: here $\beta = \sigma(\alpha)$, and  $x_1,\ldots,x_n$ (black dots) are the initial-boundary data. }
\end{figure}

We consider quad-graphs with two parallel boundaries as depicted in Figure~\ref{fig1}. On the ``left'' boundary (fields with index $1$), we impose boundary conditions associated to $q_-=0$, while on the ``right'' boundary (fields with index $n$), we impose boundary conditions associated to $q_+=0$. Given a $3$D-consistent quad equation $Q=0$ imposed on the bulk (composed of quadrilaterals), $q_-=0$ and $q_+=0$ can be different solutions to the boundary consistency condition but under the same $\sigma$.
In order to describe the map, we denote the solution of $Q(u,\tilde{u},\hat{u},\hat{\tilde{u}},\alpha,\beta)=0$ with respect to $\hat{\tilde{u}}$ by $\hat{\tilde{u}}=F(u,\tilde{u},\hat{u},\alpha,\beta)$, and  the solution of $q_\pm(x,y,z,\alpha)=0$ with respect to $z$ by $z=f_\pm(x,y,\alpha)$.  In the simplest initial-boundary-value problem we consider here, the initial data $x_1, \dots, x_n$ and $\alpha_1, \dots, \alpha_{n-1}$  will evolve  to  $x'_1, \dots, x'_n$  and $\alpha'_1, \dots, \alpha'_{n-1}$ by one-step discrete ``time", and eventually propagate to infinity.  This corresponds to a collective move of the fields and parameters from lattice site $(n, m)$ to $(n + 1, m + 1)$ within the strip, following the notations of \eqref{lattice}.

First, we take the lattice parameters to be $\alpha$ on horizontal edges and $\sigma(\alpha)$ on vertical edges. The so-constructed maps are {\em autonomous}, as the parameters remain unchanged after one-step of evolution.
In the graph on the left we have an odd number, $n=2k+1$, of variables. The upward evolution (north-east direction in our figures) is given by
\begin{equation}\label{eqo}
\begin{cases} x_1^\prime=f_-(x_1,x_2,\alpha)\,, & \\
x_{2i+1}^\prime=F(x_{2i+1},x_{2i+2},x_{2i},\alpha,\sigma(\alpha))\,, & 1\leq i<k \,,\\
x_{n}^\prime=f_+(x_n,x_{n-1},\sigma(\alpha))\,, & \\
x_{2i}^\prime=F(x_{2i},x_{2i+1}^\prime,x_{2i-1}^\prime,\alpha,\sigma(\alpha))\,, & 1\leq i \leq k\,.
\end{cases}
\end{equation}
In the graph on the right we have an even number, $n=2k+2$, of variables. The evolution upwards is given by
\begin{equation}\label{eqe}
\begin{cases}
x_1^\prime=f_-(x_1,x_2,\alpha)\,, & \\
x_{2i+1}^\prime=F(x_{2i+1},x_{2i+2},x_{2i},\alpha,\sigma(\alpha))\,, & 1\leq i \leq  k\,, \\
x_{2i}^\prime=F(x_{2i},x_{2i+1}^\prime,x_{2i-1}^\prime,\alpha,\sigma(\alpha))\,, & 1\leq i \leq k\,,\\
x_{n}^\prime=f_+(x_n,x_{n-1}^\prime,\sigma(\alpha))\,. &
\end{cases}
\end{equation}
The inverses of these maps, \ie the downward evolution (south-west direction) can be written down in a similar fashion%  using the solutions of $Q(u,\tilde{u},\hat{u},\hat{\tilde{u}},\alpha,\beta)=0$ and $q(u,\tilde{u},\hat{\tilde{u}},\alpha)=0$ with respect to
% $u$
.

We can consider similar initial-boundary value problems, but with general lattice parameters along the staircase. Here the maps need to be accompanied by an action on the parameters, and hence become {\em non-autonomous}. As this action is cyclic, one can consider the ($n-1$)-th power of this map, which is again autonomous.
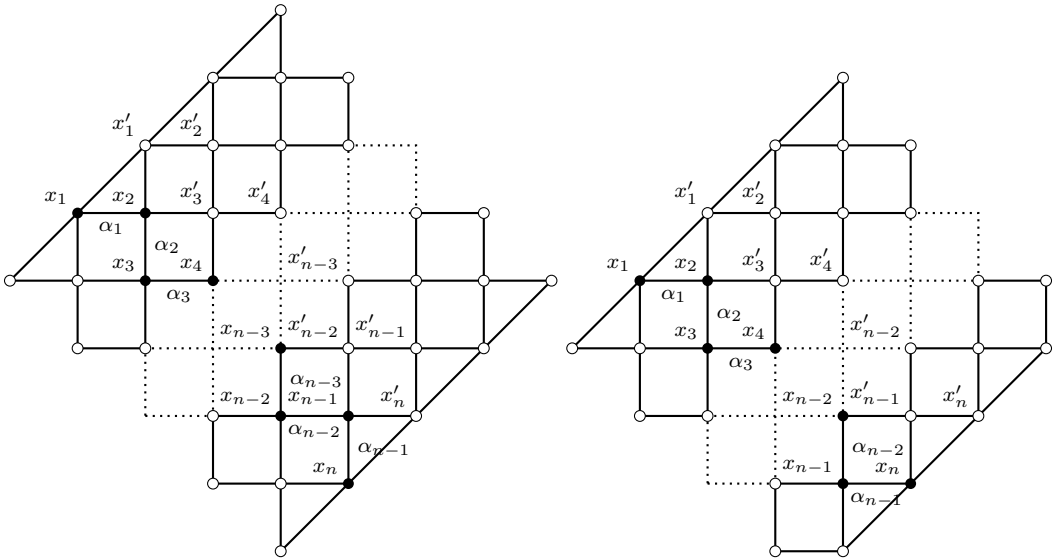
\begin{figure}[h!]
\begin{center}
\begin{tikzpicture}[scale=.9]
\draw[thick]
    (0,0)--(4,4)--(4,1)--(1,1)--(1,-1)--(2,-1)--(2,2)--(5,2)--(5,3)
    --(5,3)--(3,3)--(3,0)--(0,0)
    (4,-4)--(8,0)--(5,0)--(5,-3)--(3,-3)--(3,-2)--(6,-2)--(6,1)--(7,1)
    --(7,-1)--(4,-1)--(4,-4);
\draw[dotted,thick]
    (2,-2)--(3,-2)--(3,0)--(5,0)--(5,2)--(6,2)--(6,1)--(4,1)--(4,-1)--(2,-1)--(2,-2);
\fill
    \foreach \x in {(1,1),(2,1),(2,0),(3,0),(4,-1),(4,-2),(5,-2),(5,-3)}
            {\x circle (0.08cm)};
\fill[color=white]
    \foreach \x in {(0,0),(1,0),(1,-1),(2,-1),(3,-2),(3,-3),(4,-3),(4,-4),(4,4),(3,3),(4,3),
    (5,3),(2,2),(3,2),(4,2),(5,2),(3,1),(4,1),(6,-2),(5,-1),(6,-1),(7,-1),(5,0),
    (6,0),(7,0),(8,0),(6,1),(7,1)}
            {\x circle (0.08cm)};
\draw
    \foreach \x in {(0,0),(1,0),(1,-1),(2,-1),(3,-2),(3,-3),(4,-3),(4,-4),(4,4),(3,3),(4,3),
    (5,3),(2,2),(3,2),(4,2),(5,2),(3,1),(4,1),(6,-2),(5,-1),(6,-1),(7,-1),(5,0),
    (6,0),(7,0),(8,0),(6,1),(7,1)}
            {\x circle (0.08cm)};
\draw   (1,1) node[above left] {\footnotesize$x_1$}
        (2,1) node[above left] {\footnotesize$x_2$}
        (2,0) node[above left] {\footnotesize$x_3$}
        (3,0) node[above left] {\footnotesize$x_4$}
        (4,-1) node[above left] {\footnotesize$x_{n-3}$}
        (4,-2) node[above left] {\footnotesize$x_{n-2}$}
        (5,-2) node[above left] {\footnotesize$x_{n-1}$}
        (5,-3) node[above left] {\footnotesize$x_{n}$};
\draw   (1.5,1) node[below] {\footnotesize$\alpha_1$}
        (2,.5) node[right] {\footnotesize$\alpha_2$}
        (2.5,0) node[below] {\footnotesize$\alpha_3$}
        (4,-1.5) node[right] {\footnotesize$\alpha_{n-3}$}
        (4.5,-2) node[below] {\footnotesize$\alpha_{n-2}$}
        (5,-2.5) node[right] {\footnotesize$\alpha_{n-1}$};
\draw   (2,2) node[above left] {\footnotesize$x_1^\prime$}
        (3,2) node[above left] {\footnotesize$x_2^\prime$}
        (3,1) node[above left] {\footnotesize$x_3^\prime$}
        (4,1) node[above left] {\footnotesize$x_4^\prime$}
        (5,0) node[above left] {\footnotesize$x_{n-3}^\prime$}
        (5,-1) node[above left] {\footnotesize$x_{n-2}^\prime$}
        (6,-1) node[above left] {\footnotesize$x_{n-1}^\prime$}
        (6,-2) node[above left] {\footnotesize$x_{n}^\prime$};
\end{tikzpicture}
\begin{tikzpicture}[scale=.9]
\draw[thick]
    (0,0)--(4,4)--(4,1)--(1,1)--(1,-1)--(2,-1)--(2,2)--(5,2)--(5,3)
    --(5,3)--(3,3)--(3,0)--(0,0)
    (4,-3)--(7,0)--(5,0)--(5,-2)--(3,-2)--(3,-3)--(4,-3)--(4,-1)--(6,-1)
    --(6,1)--(7,1)--(7,0);
\draw[dotted,thick]
    (2,-2)--(3,-2)--(3,0)--(5,0)--(5,2)--(6,2)--(6,1)--(4,1)--(4,-1)--(2,-1)--(2,-2);
\fill
    \foreach \x in {(1,1),(2,1),(2,0),(3,0),(4,-1),(4,-2),(5,-2)}
            {\x circle (0.08cm)};
\fill[color=white]
    \foreach \x in {(0,0),(1,0),(1,-1),(2,-1),(3,-2),(3,-3),(4,-3),(4,4),(3,3),(4,3),
    (5,3),(2,2),(3,2),(4,2),(5,2),(3,1),(4,1),(5,-1),(6,-1),(5,0),
    (6,0),(7,0),(6,1),(7,1)}
            {\x circle (0.08cm)};
\draw
    \foreach \x in {(0,0),(1,0),(1,-1),(2,-1),(3,-2),(3,-3),(4,-3),(4,4),(3,3),(4,3),
    (5,3),(2,2),(3,2),(4,2),(5,2),(3,1),(4,1),(5,-1),(6,-1),(5,0),
    (6,0),(7,0),(6,1),(7,1)}
            {\x circle (0.08cm)};
\draw   (1,1) node[above left] {\footnotesize$x_1$}
        (2,1) node[above left] {\footnotesize$x_2$}
        (2,0) node[above left] {\footnotesize$x_3$}
        (3,0) node[above left] {\footnotesize$x_4$}
        (4,-1) node[above left] {\footnotesize$x_{n-2}$}
        (4,-2) node[above left] {\footnotesize$x_{n-1}$}
        (5,-2) node[above left] {\footnotesize$x_{n}$};
\draw   (1.5,1) node[below] {\footnotesize$\alpha_1$}
        (2,.5) node[right] {\footnotesize$\alpha_2$}
        (2.5,0) node[below] {\footnotesize$\alpha_3$}
        (4,-1.5) node[right] {\footnotesize$\alpha_{n-2}$}
        (4.5,-2) node[below] {\footnotesize$\alpha_{n-1}$};
\draw   (2,2) node[above left] {\footnotesize$x_1^\prime$}
        (3,2) node[above left] {\footnotesize$x_2^\prime$}
        (3,1) node[above left] {\footnotesize$x_3^\prime$}
        (4,1) node[above left] {\footnotesize$x_4^\prime$}
        (5,0) node[above left] {\footnotesize$x_{n-2}^\prime$}
        (5,-1) node[above left] {\footnotesize$x_{n-1}^\prime$}
        (6,-1) node[above left] {\footnotesize$x_{n}^\prime$};
\end{tikzpicture}
\end{center}
\caption{\label{fig2} More general case with $n-1$ lattice parameters.}\label{GL}
\end{figure}
With general lattice parameters (see Figure~\ref{GL}), we have the following maps. The two-dimensional map is the same as (\ref{eqe}). The three-dimensional map
is
\begin{equation}\label{eql3}
\begin{cases}
x_1^\prime=f_-(x_1,x_2,\alpha_1)\,, & \\
x_{3}^\prime=f_+(x_3,x_{2},\alpha_2)\,, & \\
x_{2}^\prime=F(x_{2},x_{3}^\prime,x_{1}^\prime,
\sigma(\alpha_2),\sigma(\alpha_1))\,, &
\end{cases}
\end{equation}
with
\begin{equation}
(\alpha_1,\alpha_2)\rightarrow (\sigma(\alpha_2),\sigma(\alpha_1))\,.
\end{equation}
The odd-dimensional map with $n=2k+1$, $k>1$ is
\begin{equation}\label{eqog}
\begin{cases} x_1^\prime=f_-(x_1,x_2,\alpha_1)\,, & \\
x_{2i+1}^\prime=F(x_{2i+1},x_{2i+2},x_{2i},\alpha_{2i+1},\alpha_{2i})\,, & 1\leq i<k-1\,, \\
x_{n}^\prime=f_+(x_n,x_{n-1},\alpha_{2k})\,, & \\
x_{2}^\prime=F(x_{2},x_{3}^\prime,x_{1}^\prime,
\alpha_{3},\sigma(\alpha_{1}))\,, & \\
x_{2i}^\prime=F(x_{2i},x_{2i+1}^\prime,x_{2i-1}^\prime,
\alpha_{2i+1},\alpha_{2i-2})\,, & 1 < i < k\,,\\
x_{n-1}^\prime=F(x_{n-1},x_{n}^\prime,x_{n-2}^\prime,
\sigma(\alpha_{2k}),\alpha_{2k-2)})\,, &
\end{cases}
\end{equation}
with
\begin{equation}
  \label{eq:oddpara}
\begin{cases}
\alpha_2\rightarrow \alpha_2^\prime=\sigma(\alpha_1)\,, & \\
\alpha_{2i-1}\rightarrow \alpha_{2i-1}^\prime =\alpha_{2i+1}\,, & 1\leq i < k\,, \\
\alpha_{2i+2}\rightarrow \alpha_{2i+2}^\prime =\alpha_{2i}\,,& 1\leq i < k\,, \\
\alpha_{2k-1}\rightarrow \alpha_{2k-1}^\prime =\sigma(\alpha_{2k})\,.
\end{cases}
\end{equation}
The even-dimensional map with $n=2k+2$, $k>0$ is
\begin{equation}\label{eqeg}
\begin{cases}
x_1^\prime=f_-(x_1,x_2,\alpha_1)\,, & \\
x_{2i+1}^\prime=F(x_{2i+1},x_{2i+2},x_{2i},\alpha_{2i+1},
\alpha_{2i})\,, & 1\leq i \leq k\,, \\
x_2^\prime=F(x_2,x_3,x_1,\alpha_3,\sigma(\alpha_1))\,, & \\
x_{2i}^\prime=F(x_{2i},x_{2i+1}^\prime,x_{2i-1}^\prime,
\alpha_{2i+1},\alpha_{2i-2})\,, & 1< i \leq k\,,\\
x_{n}^\prime=f_+(x_n,x_{n-1}^\prime,\alpha_{2k}))\,, &
\end{cases}
\end{equation}
with
\begin{equation}
\begin{cases}
\alpha_2\rightarrow \alpha_2^\prime = \sigma(\alpha_1)\,, & \\
\alpha_{2i-1}\rightarrow \alpha_{2i-1}^\prime= \alpha_{2i+1}\,, & 1\leq i < k\,, \\
\alpha_{2i+2}\rightarrow \alpha_{2i+2}^\prime = \alpha_{2i}\,, & 1\leq i < k\,, \\
\alpha_{2k+1}\rightarrow \alpha_{2k+1}^\prime =\sigma(\alpha_{2k})\,.
\end{cases}
\end{equation}
Of course, the case with general lattice parameters contains the situation of Figure~\ref{fig1} as a particular case where $\alpha_{2i+1}=\alpha$ and $\alpha_{2i}=\sigma(\alpha)$.

\subsection{Generating function for the invariants}\label{sec:genf}
Consider the maps defined above with general lattice parameters (see Figure~\ref{fig2}). Let us first recall the bulk monodromy matrix $ { T}(\lambda)$ from $x_1$ to $x_n$ as the following ordered product of Lax matrices $L$ associated to the bulk equation $Q=0$
\begin{equation}
  \label{eq:7}
  { T}(\lambda)  =  \overset{\curvearrowleft}{\prod_{j =1,\cdots,n-1}} L(x_{j+1},x_j;\alpha_j,\lambda) =L_{n,n-1}(\lambda) \dots L_{2,1}(\lambda)\,.
\end{equation}
Here, the notation $ L_{j+1,j}(\lambda)=L(x_{j+1},x_j;\alpha_j,\lambda)$ is understood. Note that $T(\lambda)$ depends on all the $x_j$'s and $\alpha_j$'s but we do not show this dependence
explicitly for conciseness. The updated values of $T(\lambda)$, \ie its value at $x_j^\prime$ and $\alpha_j^\prime$, will be simply denoted by $T^\prime(\lambda)$.  Similarly, let $\check{T}$ denote the reverse-ordered monodromy matrix from $x_n$ to $x_1$
\begin{equation}
  \label{eq:7}
  \check{ T}(\lambda)  =  \overset{\curvearrowright}{\prod_{j =1,\cdots,n-1}} L(x_{j},x_{j+1};\alpha_j,\lambda)  =L_{1,2}(\lambda) \dots L_{n-1,n}(\lambda) \,.
\end{equation}
Now let $n=2k+2$ for even $n$, and $n=2k+1$ for odd $n$.   Inspired by Sklyanin's construction \cite{sklyanin1987boundary,sklyanin1988boundary}, we now define the so-called double-row monodromy matrix
${\cal T}(\lambda)$ in the form
\begin{equation}
  \label{eq:7}
  {\cal T}(\lambda)  = K_-(x_1;\sigma(\lambda))\, \check{T}(\sigma(\lambda))\,K_+(x_n;\lambda)\,T(\lambda)\,,
\end{equation}
where $ K_-$  satisfies
\begin{align}\label{eq:bzc1}
  L_{1', 2}( \lambda)  L_{2, 1}( \lambda) K_{-}(x_1, \sla) &= \epsilon_- K_{-}(x'_{1},\sla) L_{1', 2}(\sla)  L_{2, 1}(\sla)\,,\quad \epsilon_-=\pm 1\,,
\end{align}
corresponding to $q_-(x_1,x_2,x_1';\alpha_1) = 0$, and
$ K_+$  satisfies
\begin{align}
\label{eq:bzc2} L_{n', n-1}( \sla)  L_{n-1, n}( \sla) K_{+}(x_n,\lambda) &= \epsilon_+ K_{+}(x'_n,\lambda) L_{n', n-1}(\lambda)  L_{n-1, n}(\lambda)\,,\quad \epsilon_+=\pm 1\,,
\end{align}
corresponding to $q_+(x_n,x_{n-1},x'_{n};\alpha_{2k}) = 0$ if $n$ is odd, \ie $n=2k+1$ or
\begin{align}
\label{eq:bzc3} L_{n', n-1'}( \sla)  L_{n-1', n}( \sla) K_{+}(x_n,\lambda) &= \epsilon_+ K_{+}(x'_n,\lambda) L_{n', n-1'}(\lambda)  L_{n-1', n}(\lambda)\,,\quad \epsilon_+=\pm 1\,,
\end{align}
corresponding to $q_+(x_n,x'_{n-1},x'_{n};\alpha_{2k}) = 0$ if $n$ is even, \ie $n=2k+2$.
We have the following result.
\begin{theorem}\label{th:double_row}
 Let $L$ be the Lax matrix associated to $Q =0$  from the ABS list, and $K_-$ and $K_+$ be the boundary matrices associated to the integrable boundary equation $q_-=0$ and $q_+=0$, normalised as explained in Section \ref{bd_ZC} \ie such that the boundary zero curvature conditions \eqref{eq:bzc1} and \eqref{eq:bzc2} (for odd  $n=2k+1$) or \eqref{eq:bzc3} (for even $n=2k+2$) hold. Then, the monodromy matrix ${\cal T}(\lambda)$ defined in \eqref{eq:7} and its updated version $ {\cal T}^\prime(\lambda)$  are related by\footnote{We stress that $\ell(\alpha,\lambda)$ is here the normalisation of $L$ taking $\eqref{rescaleL1}$ into account.}
\begin{equation}
\label{TTprime}
{\cal T}^\prime(\lambda)\E =\epsilon_-\,\epsilon_+\,\frac{\ell(\sigma(\alpha_1),\lambda)
\ell(\sigma(\alpha_{2k}),\sigma(\lambda))}{\ell(\alpha_1,\sigma(\lambda))\ell(\alpha_{2k},\lambda)}
\E\,{\cal T}(\lambda)\,,
\end{equation}
where
\begin{equation}
\E =L(x_1^\prime,x_2;\sigma(\alpha_1),\lambda)L(x_2,x_1;\alpha_1,\lambda).
\end{equation}
\end{theorem}
{\bf Proof:} Let us consider the odd case $n=2k+1$. The even case $n = 2k +2$ is completely analogous. For simplicity, we use the notations
$L_{j+1,j}(\lambda)=L(x_{j+1},x_j;\alpha_j,\lambda) $, $L_{j+1,j'}(\lambda  )=L(x_{ j+1}, x'_{ j}; \alpha_j , \lambda)  $, etc., by dropping the dependence of the lattice parameter as it is always associated to the edge connecting the two adjacent vertices.
%We also use $\eta$ (resp.~$\beta$) to denote $\sigma(\lambda)$ (resp.~$\sigma(\alpha)$).
%The map (\ref{eqog}) is the solution of the set of equations
%\[
%\begin{cases}
%q(x_1,x_2,x_1^\prime,\alpha_1)=0, q(x_n,x_{n-1},x_n^\prime,\alpha_{n-1})=0&\\
%Q(x_{2i+1},x_{2i+2},x_{2i},x_{2i+1}^\prime,\alpha_{2i+1},\alpha_{2i})=0& 1\leq i < k\\
%Q(x_{2i},x_{2i+1}^\prime,x_{2i-1}^\prime,x_{2i}^\prime,
%\alpha_{2i+1},\alpha_{2i})=0& 1\leq i \leq k.
%\end{cases}
%\]
It follows from \eqref{eq:invL} that
\begin{equation}
T(\lambda) =\gamma\, L_{2k+1,2k}(\lambda)\,L_{2k,2k+1^\prime}(\lambda)\,L_{2k+1^\prime,2k}(\lambda)
\left(\overset{\curvearrowleft}{\prod_{j =2,\cdots,2k-1}} L_{j+1,j}(\lambda)\right)L_{2,1^\prime}(\lambda)\,L_{1^\prime,2}(\lambda)\,L_{2,1}(\lambda)\,,
\end{equation}
where
\begin{equation}
\gamma=\frac{1}{\ell(\sigma(\alpha_{1}),\lambda)\ell(\sigma(\alpha_{2k}),\lambda)}  \,.
\end{equation}
Using the bulk zero curvature conditions
\begin{equation}
L_{2j,2j-1}(\lambda)\,L_{2j-1,2j-2}(\lambda)=L_{2j,2j-1^\prime}(\lambda)\,L_{2j-1^\prime,2j-2}(\lambda)\,,
\end{equation}
for $j=2,\dots,k$, and then
\begin{equation}
L_{2j+1^\prime,2j}(\lambda)\,L_{2j,2j-1^\prime}(\lambda)=L_{2j+1^\prime,2j^\prime}(\lambda)\,L_{2j^\prime,2j-1^\prime}(\lambda)\,,
\end{equation}
for $j=1,\dots,k$, we find
\begin{equation}
T(\lambda)=\gamma\,L_{2k+1,2k}(\lambda)\,L_{2k,2k+1^\prime}(\lambda)\,T^\prime(\lambda)\,L_{1^\prime,2}(\lambda)\,L_{2,1}(\lambda)\,.
\end{equation}
Similarly for $\check{T}(\eta)$, one has
\begin{equation}
\check{T}(\sigma(\lambda)) = \check{\gamma}\,L_{1,2}(\sigma(\lambda))\,L_{2,1'}(\sigma(\lambda))\,\check{T}' (\sigma(\lambda))\, L_{2k+1',2k}(\sigma(\lambda))\,L_{2k,2k+1}(\sigma(\lambda))\,,
\end{equation}
where
\begin{equation}
\check{\gamma} =\frac{1}{\ell(\sigma(\alpha_{1}),\sigma(\lambda))\ell(\sigma(\alpha_{2k}),\sigma(\lambda))}
\end{equation}
Therefore
\begin{align}
  \E{\cal T}(\lambda)  &= \gamma \check{\gamma} \left[L_{1',2}(\lambda) L_{2,1}(\lambda)K_-(x_1;\sigma(\lambda))\,L_{1,2}(\sigma(\lambda))\,L_{2,1'}(\sigma(\lambda))\right]\,\check{T}'(\sigma(\lambda)) \nonumber\\
                       & \times \left[  L_{2k+1',2k}(\sigma(\lambda))\,L_{2k,2k+1}(\sigma(\lambda)) K_{+}(x_{2k+1},\lambda)  L_{2k+1,2k}(\lambda)L_{2k,2k+1'}(\lambda)\right]T'(\lambda)\, \E\,.
\end{align}
Taking \eqref{eq:bzc1} and \eqref{eq:bzc2} into account, the terms in square brackets can be reduced to
\begin{align}
&  L_{1',2}(\lambda) L_{2,1}(\lambda)K_-(x_1;\sigma(\lambda))\,L_{1,2}(\sigma(\lambda))\,L_{2,1'}(\sigma(\lambda)) = \ell(\alpha_1,\sigma(\lambda))\ell(\sigma(\alpha_1),\sigma(\lambda))\epsilon_-K_-(x'_1;\sigma(\lambda))\,, \\
&  L_{2k+1',2k}(\sigma(\lambda))\,L_{2k,2k+1}(\sigma(\lambda)) K_{+}(x_{2k+1},\lambda)  L_{2k+1,2k}(\lambda)L_{2k,2k+1'}(\lambda)   = \ell(\sigma(\alpha_{2k}),\lambda)\ell(\alpha_{2k},\lambda) \epsilon_+ K_{+}(x'_{2k+1},\lambda)\,.
\end{align}
Matching all the factors completes the proof. \finprf
\medskip

\begin{corollary}\label{cor:1}
In the case of autonomous maps (\ref{eqo}-\ref{eqe}), or, in the non-autonomous case, if $\ell(\alpha,\lambda)$ is such that $\ell(\alpha,\sigma(\lambda))/\ell(\sigma(\alpha),\lambda)=1$, then function ${\mathfrak t}(\lambda)=\text{Tr}\,{\cal T}(\lambda) $ satisfies
\begin{equation}
{\mathfrak t}^\prime(\lambda) = \epsilon_-\epsilon_+\,{\mathfrak t}(\lambda)\,,
\end{equation}
and thus can be taken as a generating function for the invariants of the corresponding map. In the case $\epsilon_-\epsilon_+=1$, this is automatic. In the case $\epsilon_-\epsilon_+=-1$, a quantity $I$ extracted from ${\mathfrak t}(\lambda)$ is a $2$-integral. To obtain an invariant, it suffices to take $I^2$ for instance.
\end{corollary}
{\bf Proof:}
In the autonomous case, we have $\alpha_1=\alpha=\sigma(\alpha_{2k})$, hence $\frac{\ell(\sigma(\alpha_1),\lambda)
	\ell(\sigma(\alpha_{2k}),\sigma(\lambda))}{\ell(\alpha_1,\sigma(\lambda))\ell(\alpha_{2k},\lambda)}=1$. The latter is also true in the non-autonomous case if $\ell(\alpha,\sigma(\lambda))/\ell(\sigma(\alpha),\lambda)=1$. Therefore, in both cases \eqref{TTprime} reduces to
\begin{equation}
{\cal T}^\prime(\lambda)\E=\epsilon_-\,\epsilon_+\,\E\,{\cal T}(\lambda)\,,
\end{equation}
and the result follows.
\finprf
\medskip

With general lattice parameters along the staircase, the maps (\ref{eqeg}) and (\ref{eqog}) are non-autonomous. However, the $(n-1)$-st power of these maps are autonomous, and we obtain results similar to Corollary \ref{cor:1}. The double-row monodromy matrix provides $k$-integrals for the non-autonomous maps, where $k=n-1$ or $k=2(n-1)$ when $n$ even and $\epsilon_-\epsilon_+=-1$ ($k$-integrals were introduced in \cite{HBQC}).

\begin{corollary}\label{cor:2}
Let ${\cal T}^{(n-1)}$ denote the monodromy matrix as defined in \eqref{eq:7} after the $(n-1)$-st iterate of the map. Then ${\cal T}$ satisfies
  \begin{equation}\label{eq:nstepmap}
{\cal T}^{(n-1)} \E_n =(\epsilon_- \epsilon_+)^{n-1}\E_n {\cal T} \,,
  \end{equation}
  where
  \begin{equation}
   \E_n=\E^{(n-2)}\E^{(n-1)}\dots\E^{(1)} \E\,,
  \end{equation}
  and
  \begin{equation}
    \E^{(j)} = L(x_1^{(j+1)},x_2^{(j)};\sigma(\alpha^{(j)}_1),\lambda)L(x^{(j)}_2,x^{(j)}_1;\alpha^{(j)}_1,\lambda)\,,\quad 1\leq j \leq n-2\,,
  \end{equation}
  with the superscript $j$ denoting a $j$-step evolution of the associated fields and parameters.
\end{corollary}
{\bf Proof:}
% For simplicity, we rewrite  \eqref{TTprime} in Prop.~\ref{th:double_row} as
% \begin{equation}\label{eq:TP}
%   {\cal T}^{(1)}(\lambda)\E=\epsilon_-\,\epsilon_+\,\Sigma\,\E\,{\cal T}(\lambda)\,,\quad \Sigma =\frac{\ell(\alpha_1,\sigma(\lambda))\ell(\alpha_{2k},\lambda)}{\ell(\sigma(\alpha_1),\lambda)\ell(\sigma(\alpha_{2k}),\sigma(\lambda))}\,.
% \end{equation}
% Here $\Sigma$ is the factor appearing in right-hand side of \eqref{TTprime}.
It suffices to show that
\begin{equation}\label{eq:cancelspara}
\frac{\ell(\sigma(\alpha^{(n-2)}_1),\lambda)\ell(\sigma(\alpha^{(n-2)}_{2k}),\sigma(\lambda))}{\ell(\alpha^{(n-2)}_1,\sigma(\lambda))\ell(\alpha^{(n-2)}_{2k},\lambda)}
\frac{\ell(\sigma(\alpha^{(n-1)}_1),\lambda)\ell(\sigma(\alpha^{(n-1)}_{2k}),\sigma(\lambda))}{\ell(\alpha^{(n-1)}_1,\sigma(\lambda))\ell(\alpha^{(n-1)}_{2k},\lambda)}
\dots \frac{\ell(\sigma(\alpha_1),\lambda)\ell(\sigma(\alpha_{2k}),\sigma(\lambda))}{\ell(\alpha_1,\sigma(\lambda))\ell(\alpha_{2k},\lambda)} = 1\,.
\end{equation}
Then \eqref{eq:nstepmap} follows directly from \eqref{TTprime} and its updates.
The above equality involves parameters $\alpha_1, \alpha_{2k}$ and their updates.
Let us consider the odd case $n = 2k+1$. There are $2k$ parameters along the staircase, namely, $\alpha_1, \alpha_2, \dots, \alpha_{2k}$. One could make the following identification between  $\alpha_1, \alpha_{2k}$ and their updates:
\begin{align*}
 & \alpha^{(1)}_1 = \alpha_{3}\,,~ \alpha^{(2)}_1 = \alpha_{5}\,,~ \dots~ \,,\alpha_1^{(k-1)} =\alpha_{2k-1}\,,~ \alpha_1^{(k)} =\sigma(\alpha_{2k})\,, ~ \alpha_1^{(k+1)} =\sigma(\alpha_{2k-2}) \,, ~\dots~\,, \alpha_1^{(2k-1)} =\sigma(\alpha_{2})  \,,\\
 &  \alpha^{(1)}_{2k} = \alpha_{2k-2}\,,~ \alpha^{(2)}_{2k} = \alpha_{2k-4}\,,~ \dots~\,, \alpha_{2k}^{(k-1)} =\alpha_{2}\,,~ \alpha_{2k}^{(k)} =\sigma(\alpha_{1})\,, ~ \alpha_{2k}^{(k+1)} =\sigma(\alpha_{3}) \,, ~\dots~\,, \alpha_{2k}^{(2k-1)} =\sigma(\alpha_{2k-1})\,.
\end{align*}
The equality \eqref{eq:cancelspara} follows from the above identifications.  The even case can be proved in a  similar way.
\finprf
\medskip

In general, the operation of extracting invariants from ${\mathfrak t}(\lambda)$ has to be done carefully as it assumes that there is a natural way to expand ${\mathfrak t}(\lambda)$ in $\lambda$. The normalization of $L$ and $K$ as well as their dependence on $\lambda$ play a role as they could lead to ${\mathfrak t}(\lambda)$ not being a Laurent polynomial in $\lambda$ for instance (which is the simplest case for extraction). We will illustrate the procedure on examples for which this can be done relatively easily.

\section{Examples of open boundary reductions for the H1 and Q1($\delta=0$) equations}\label{examples}
Following the construction of open boundary reductions on a strip of the $\ZZ^2$-lattice, we provide some explicit maps of dimension $n \leq 4$ for the H1 and Q1($\delta=0$) equations from the ABS list (see Section~\ref{sec:ex} for their explicit forms and their Lax and boundary matrices). 
The invariants of these maps are obtained by taking the trace of the double-row monodromy matrix defined in~\eqref{eq:7}. Each map we compute here possesses enough invariants, which suggest that they are integrable.
\subsection{H1 additive case with two different boundary equations}
Consider the case where $q_-$ and $q_+$ are different. We use
\begin{equation} \label{qus}
q_-(x,y,z;\alpha)=x+z\,,\quad q_+(x,y,z;\alpha)=y \left( z-x \right) +\alpha-\mu\,,
\end{equation}
with  $\sigma(\alpha)=-\alpha+2\mu$ (see Table~\ref{Tabh1}). First we consider the maps (\ref{eqo}), (\ref{eqe}) with lattice parameters $\alpha,\sigma(\alpha)$. However, a more convenient parameter is $c=2(\mu-\alpha)$.

The $2$-dimensional map (\ref{eqe})
\begin{equation}
(x_1,x_2)\mapsto\left(-x_1,x_{{2}}+{\frac {c}{2\,x_{{1}}}}\right)
\end{equation}
is an involution. 

The 3-dimensional map (\ref{eqo}) is given by
\begin{equation}
\label{map1}
(x_1,x_2,x_3)\mapsto\left(-x_1,
x_{{2}}+{\frac {2cx_{{2}}}{2\,x_{{1}}x_{{2}}+2\,x_{{2}}x_{{3}}-c}},
x_{{3}}-{\frac {c}{2x_{{2}}}}\right)\,.
\end{equation}
There is an obvious invariant given by $x_1^2$. Another independent invariant is obtained using our construction. Using
\begin{equation}
\label{choiceKL}
L(x,y;\alpha,\lambda)=\frac{1}{\sqrt{\lambda-\mu}}
\bma
 y & \alpha-\lambda-xy\\
1 & -x
 \ema\,,\quad K_-(x,\lambda)=\bma 1 & \frac{\mu-\lambda}{x} \\ 0 & 1\ema \,,\quad K_+(x,\lambda)=\bma -1 & 0 \\ 0 & 1\ema\,,
\end{equation}
we find
\begin{equation}
\mathfrak{t}(\lambda)=\frac{I}{\lambda-\mu}\,,\quad I=\frac{(x_1+x_3)(2x_2(x_1-x_3)+c)}{x_1}\,.
\end{equation}
Recall that $\epsilon_-=-1=-\epsilon_+$ here, hence we know that $I_1'=-I_1$, which indeed can be checked directly. This means that $I_1$ is a $2$-integral. In order to obtain an invariant of the map, one can multiply $I_1$ by $x_1$ which is also a $2$-integral. In particular, we do not need to consider $I_1^2$ for instance. Therefore, we have the following two invariants for the map \eqref{map1}
\begin{equation}
x_1^2\,,\quad (x_1+x_3)(2x_2(x_1-x_3)+c)\,.
\end{equation}
In terms of variables $y_{{1}}=x_{{3}}$, $y_{{2}}=x_{{3}}-{\frac {c}{2\,x_{{2}}}}$, $y_{{3}}=x_{{1}}$, the map reads
\begin{equation} \label{iop}
\gamma:(y_1,y_2,y_3)\mapsto\left(y_2,
y_{{3}}-{\frac { \left( y_{{1}}-y_{{2}} \right)  \left( y_{{3}}+y
_{{2}} \right) }{2\,y_{{1}}}}
,
-y_3\right),
\end{equation}
and the preserved integral is
\begin{equation}
J(y)={\frac { \left( y_{{2}}-y_{{3}} \right)  \left( y_{{1}}+y_{{3}}
		\right) }{y_{{1}}-y_{{2}}}}\,.
\end{equation}
The map is understood geometrically as $\gamma=\mathfrak{s}\circ\iota$ where $\mathfrak{s}:(y_1,y_2,y_3)\rightarrow(y_2,y_1,-y_3)$ is an anti-symmetry switch and $\iota:(y_1,y_2,y_3)\rightarrow(y_1^\prime,y_2,y_3)$ is an anti-horizontal switch, cf. \cite{Dui} , \ie
\begin{equation}
J(\mathfrak{s}(y))=-J(y), \quad J(\iota(y))=-J(y)\,.
\end{equation}
For fixed $y_3$ each curve $J(y)=j$ intersects horizontal (and vertical) lines once, $\iota$ maps $y$ on $J(y)=j$ to the unique point $y^\prime$ on $J(y)=-j$ that has the same $y_2,y_3$, and $\iota$ being an anti-symmetry switch means that the reflection in the line $y_1=y_2$ of the line $J(y)=j$ with $y_3$ fixed
equals the line $J(y)=-j$ at $-y_3$.

The four dimensional map equals
\begin{equation}
(x_1,x_2,x_3,x_4)\rightarrow \left(-x_1,
x_{{2}}+{\frac {c \left( x_{{2}}-x_{{4}} \right) }{ \left( x_{{2}}-x_{
{4}} \right)  \left( x_{{1}}+x_{{3}} \right) -c}},
x_3+\frac{c}{x_4-x_2},
x_4+\frac{c(x_2-x_4)}{2x_3(x_4-x_2)+2c}\right)\,.
\end{equation}
With \eqref{choiceKL}, we find
\begin{equation}
	\mathfrak{t}(\lambda)=\frac{I}{(\lambda-\mu)^2}\,,\quad I=\frac{(x_1+x_3)(2x_3(x_4-x_2)+c)((x_2-x_4)(x_1-x_3)+c)}{x_1}\,.
\end{equation}
The situation is similar to the $3$-dimensional case. We have an obvious invariant $x_1^2$ and another one easily constructed from $I$, which satisfies $I'=-I$ (since $\epsilon_-\epsilon_+=-1$), by multiplying it by $x_1$.
In terms of the variables $y_1=x_3$, $y_2=\frac{x_3(x_2-x_4)+c}{x_2-x_4}$, $y_3=x_1$, the map can be written as $\mathfrak{s}\circ\iota$, where $\mathfrak{s}(y)=(y_2,y_1,-y_3)$ and $\iota(y)=(y_1^\prime,y_2,y_3)$ with
\begin{equation}
y_1^\prime=y_2\left(1+\frac{4\,y_2(y_1+y_3)}{3y_1y_2+y_1y_3-y_2^2+y_2y_3}\right)
\end{equation}
are two involutions which leave
\begin{equation}
{\frac { \left( y_{{2}}-y_{{3}} \right)  \left( y_{{1}}+y_{{3}}
 \right)  \left( y_{{1}}+y_{{2}} \right) }{ \left( y_{{1}}-y_{{2}}
 \right) ^{2}}}
\end{equation}
invariant.

Let us now consider the non-autonomous $3$-dimensional map (\ref{eql3}) with general parameters, with $q_-=0$ on the left boundary and $q_+=0$ on the right boundary as before. In that case, the map reads
\begin{equation}
(x_1,x_2,x_3;\alpha_1,\alpha_2)\mapsto\left(-x_1,
x_2+{\frac {x_2(\alpha_{{2}}-\alpha_{{1}})}{x_{{2}} \left( x_{{1}}+x_{{3}}
 \right) +\mu-\alpha_{{2}}}},
 x_3+\frac{\mu-\alpha_2}{x_2};\sigma(\alpha_2),\sigma(\alpha_1)\right).
\end{equation}
With \eqref{choiceKL}, we have
\begin{equation}
L(x,y,\alpha,\lambda)L(y,x,\alpha,\lambda)=\ell(\alpha,\lambda)\id \,,\quad \ell(\alpha,\lambda)=\frac{\alpha-\lambda}{\lambda-\mu}\,.
\end{equation}
Hence, here $\ell(\alpha,\lambda)$ satisfies
\begin{equation}
\frac{\ell(\alpha,\sigma(\lambda))}{\ell(\sigma(\alpha),\lambda)}=1\,,
\end{equation}
thus ensuring that the ratio in \eqref{TTprime} is one. As in the autonomous case we find that $\mathfrak{t}(\lambda)$ provides us with a $2$-integral:
\begin{equation}
\mathfrak{t}(\lambda)=\frac{I}{2(\lambda-\mu)}\,,\quad I=\frac{x_{{2}}({x_{{1}}}^{2}-{x_{{3}}}^{2})+\mu(x_{{1}}-\,x_{{3}})-
	\alpha_{{1}}x_{{1}}+\alpha_{{2}}x_{{3}}}{x_1}\,,
\end{equation}
and a direct calculation gives $I'=-I$. With the same reasoning as before, we get the following two invariants
\[
x_1^2\,,\quad x_{{2}}({x_{{1}}}^{2}-{x_{{3}}}^{2})+\mu(x_{{1}}-\,x_{{3}})-
\alpha_{{1}}x_{{1}}+\alpha_{{2}}x_{{3}}.
\]
The square of this map leaves $x_1,\alpha_1,\alpha_2$ invariant, and its action on $x_2,x_3$ is
\begin{equation}
\bma
x_2 \\
x_3
\ema
\mapsto
\bma
{\frac { \left( x_{{1}}x_{{2}}+x_{{2}}x_{{3}}+\mu-\alpha_{{1}}
		\right)  \left( {x_{{1}}}^{2}{x_{{2}}}^{2}-{x_{{2}}}^{2}{x_{{3}}}^{2}
		+\mu\,x_{{1}}x_{{2}}-\mu\,x_{{2}}x_{{3}}-\alpha_{{1}}x_{{1}}x_{{2}}+
		\alpha_{{1}}x_{{2}}x_{{3}}+\mu\,\alpha_{{1}}-\mu\,\alpha_{{2}}-\alpha_
		{{1}}\alpha_{{2}}+{\alpha_{{2}}}^{2} \right) }{ \left( {x_{{1}}}^{2}x_
		{{2}}-x_{{2}}{x_{{3}}}^{2}+\mu\,x_{{1}}-\mu\,x_{{3}}-2\,\alpha_{{1}}x_
		{{1}}+\alpha_{{2}}x_{{1}}+\alpha_{{2}}x_{{3}} \right)  \left( x_{{1}}x
		_{{2}}+x_{{2}}x_{{3}}+\mu-\alpha_{{2}} \right) }} \\
	x_{{3}}+{\frac { \left( x_{{1}}+x_{{3}} \right)  \left( \alpha_{{1}}-
			\alpha_{{2}} \right) }{x_{{1}}x_{{2}}+x_{{2}}x_{{3}}+\mu-\alpha_{{1}}}}
\ema\,.
\end{equation}
In terms of variables $y_1=x_3$, $y_2=x_3+{\frac { \left( x_{{1}}+x_{{3}} \right)  \left( \alpha_{{1}}-\alpha_{{
2}} \right) }{x_{{1}}x_{{2}}+x_{{2}}x_{{3}}+\mu-\alpha_{{1}}}}
$ this map reads
\begin{equation} \label{ipo}
\delta:(y_1,y_2)\mapsto\left(y_{{2}},-{\frac {{x_{{1}}}^{2}y_{{1}}-2\,{x_{{1}}}^{2}y_{{2}}+y_{{1}}
{y_{{2}}}^{2}}{{x_{{1}}}^{2}-2\,y_{{1}}y_{{2}}+{y_{{2}}}^{2}}}\right)
\end{equation}
which leaves invariant the ratio
\begin{equation}
R={\frac {{x_{{1}}}^{2}-y_{{1}}y_{{2}}}{y_{{1}}-y_{{2}}}}\,.
\end{equation}
The map (\ref{ipo}) is similar to the map (\ref{iop}), it can be written as $\delta=\mathfrak{s}\circ\iota$ where $\mathfrak{s}:(y_1,y_2)\rightarrow(y_2,y_1)$ is an anti-symmetry switch for $R$ and $\iota:(y_1,y_2)\rightarrow(y_1^\prime,y_2)$ is the anti-horizontal switch. We note that the degree growth of these maps is linear, which indicates that they are linearisable.

Invariants can be calculated using computer algebra for $n$-dimensional maps with $n\leq 7$ quite easily. For H1 with two different boundary equations, as in (\ref{qus}),  we found $\lfloor (n+1)/2\rfloor$  functionally independent integrals.

\subsection{Q1($\delta =0$) multiplicative case with two different boundary equations}
Consider the Q1($\delta =0$) equation in the form \eqref{eq:q102} with two different boundary equations taken from Table~\ref{Tabq2}. We use the following $q_\pm$ under the multiplicative involution $\sigma(\alpha)=\frac{\mu^2}{\alpha}$:
\begin{equation}
  q_-(x,y,z,\alpha)=\alpha^2(x-y)+\mu^2(y-z)\,,\quad q_+(x,y,z,\alpha)=\alpha^2(x-y)z+\mu^2(y-z)x\,.
\end{equation}
We use the general lattice parameters in this example. The $2$-dimensional map reads
\begin{equation}
(x_1,x_2)\mapsto\left(\frac{x_1 + (-1 + c^2) x_2}{c^2} , x_2 \frac{(x_1 + (-1 + c^2) x_2)}{c^2 x_1 }\right)\,,
\end{equation}
where $c=\frac{\mu}{\alpha}$. An $N$-step iteration of the map yields
\begin{equation}
(x_1,x_2)\mapsto\left(x_1y^{N} ,x_2 y^{N}\right)  \,,\quad y  =\frac{x_1 + ( c^2-1 ) x_2}{c^2x_1}\,.
\end{equation}
By taking the trace of the monodromy matrix, one obtains one invariant $x_1/x_2$ which can be easily checked by the above general expression of the maps.

The $3$-dimensional map with generic parameters $\alpha_1,\alpha_2$ is
\begin{equation}
  (x_1,x_2, x_3)\mapsto\left(c_1^2 (x_1 - x_2) + x_2,
x_2\frac{x_2^2 + c_1^2 (x_1 - x_2) (x_2 - x_3) + x_1 x_3 - 2 x_2 x_3}{x_2^2 + c_2^2 (x_1 - x_2) (x_2 - x_3) + x_1 x_3 - 2 x_2 x_3},
\frac{ x_2 x_3}{c_2^2  (x_2 - x_3)  +x_3}
\right)\,,
\end{equation}
where $c_j = \frac{\alpha_j}{\mu}$, $j=1,2$ and the change of parameters $(\alpha_1,\alpha_2)\mapsto  (\sigma(\alpha_2),\sigma(\alpha_1))$, \ie
$(c_1,c_2)\mapsto  (1/c_2,1/c_1)$, 
is understood. The boundary matrices are given in Table \ref{Tabq2},
\begin{equation}
K_-(x,\lambda)=\bma -\frac{\lambda}{\mu} & \frac{(\lambda^2+\mu^2)x}{\lambda\mu} \\ 0 & \frac{\mu}{\lambda} \ema\,,\quad
K_+(x,\lambda)=\bma -\frac{\mu}{\lambda} & 0 \\ -\frac{\lambda^2+\mu^2}{\lambda\mu x} & \frac{\lambda}{\mu} \ema\,.
\end{equation}
The Lax matrix \eqref{LQ1} with scaling \eqref{rescaleL1} yields
\begin{equation}
L(x,y;\alpha,\lambda)=\frac{\mu}{\alpha^2(x-y)\sqrt{\lambda^2-\mu^2}}\bma \alpha^2(x-y) -\lambda^2 x& \lambda^2 xy \\   -\lambda^2 & \alpha^2(x-y)+ \lambda^2 y\ema\,,
\end{equation}
which satisfies
$L(x,y;\alpha,\lambda)L(y,x;\alpha,\lambda)=\ell(\alpha,\lambda)\id $ with $\ell(\alpha,\lambda)=\frac{\mu^2(\alpha^2-\lambda^2)}{\alpha^2(\lambda^2-\mu^2)}$. This implies that
\begin{equation}
\frac{\ell(\alpha,\sigma(\lambda))}{\ell(\sigma(\alpha),\lambda)}=\frac{\mu^2}{\alpha^2}\,,
\end{equation}
and hence that the ratio in \eqref{TTprime} is non-trivial. In this example, it is possible to further re-scale $L\rightarrow\zeta(\alpha)L={\cal L}$, by a function of $\alpha$ only, without changing the bulk zero curvature or the boundary zero curvature equations. We introduce
\begin{equation}
{\cal L}(x,y;\alpha,\lambda)=\sqrt{\alpha}L(x,y;\alpha,\lambda)\,.
\end{equation}
This gives ${\cal L}(x,y;\alpha,\lambda){\cal L}(y,x;\alpha,\lambda)={\ell^\ast}(\alpha,\lambda)\id $ where ${\ell^\ast}(\alpha,\lambda)=\frac{\mu^2(\alpha^2-\lambda^2)}{\alpha(\lambda^2-\mu^2)}$ now satisfies 
\begin{equation}
\frac{\ell^\ast(\alpha,\sigma(\lambda))}{\ell^\ast(\sigma(\alpha),\lambda)}=1\,,
\end{equation}
as desired. Equipped with this ${\cal L}$ and $K_\pm$, the trace of the double-row monodromy matrix is invariant (recall that $\epsilon_-\epsilon_+=1$ here). We find
\begin{equation}
\mathfrak{t}(\lambda)=\frac{\mu^2}{(\lambda^2-\mu^2)^2}\left((\lambda^2+\mu^2)^2 C+ 2\lambda^2\mu^2\frac{(c_1^2 + 1)(c_2^2 + 1)}{c_1c_2}\right)\,,
\end{equation}
where
\begin{equation}
C=\frac{c_2^2 x_1 (x_2 - x_3)^2 +
 c_1^2 (x_1 - x_2) (c_2^2 (x_1 - x_3) (x_2 - x_3) + (x_1 - x_2) x_3)}{c_1c_2 (x_1 - x_2) (x_2 - x_3) x_3}\,.
\end{equation}
The non-autonomous $3$-dimensional map can be reduced to a non-autonomous $2$-dimensional map. Using reduced variables,
$ z_1 = (x_1-x_2)/x_3$, $z_2=x_2/x_3-1$, we get
\begin{equation} \label{eq}
   (z_1,z_2,c_1,c_2)\mapsto\left( \frac{z_1 (1 + c_2^2 z_2) (c_2^2 z_2 (1 + z_2) + c_1^2 (z_1 + c_2^2 z_1 z_2))}{(1 + z_2) (z_1 + z_2 + c_2^2 z_1 z_2 + z_2^2)}, \frac{z_2 (c_2^2 z_2 (1 + z_2) + c_1^2 (z_1 + c_2^2 z_1 z_2))}{z_1 + z_2 + c_2^2 z_1 z_2 + z_2^2}, \frac{1}{c_2},\frac{1}{c_1}\right)  \,,
 \end{equation}
and a reduced invariant
 \begin{equation} \label{inv}
  C=  \frac{c_2^2 z_2^2 (1 + z_1 + z_2) + c_1^2 z_1 (z_1 + c_2^2 z_1 z_2 + c_2^2 z_2^2)}{c_2^2 z_1 z_2}\,.
 \end{equation}

\begin{figure}[h]
\begin{center}
\includegraphics[width=6cm]{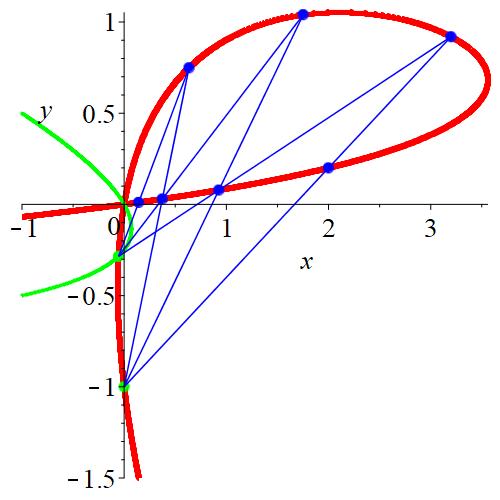}
\caption{ \label{ic} One of the involution points, $q=(-3850/69789, -154/541)$, lies on an involution curve (green), which is given by $(xy\alpha^2 + y^2)\beta^2 + (x + y)\alpha^2=0$. We have taken $\alpha=1/4,\beta=1$ and $C=337/100$.}
\end{center}
\end{figure}

The square of (\ref{eq}) is a map of the plane, which is written in terms of $x=z_1,y=z_2,\alpha=c_1^2,\beta=c_2^2$ as
\begin{equation}
\gamma:(x,y)\mapsto \frac { \left( x+y \right)  \left(
\alpha x + \beta(\alpha x + y+1)y
\right) ^{2}}{ \beta\, \left(
x+(x\beta+y+1)y
\right) \left(
(\alpha^{2}+\beta)x{y}^{2}
+\alpha(\beta{x}^{2}+{y}^{2})y
+\alpha(x+y)^2
\right) }\left({\frac {x \left(
\alpha x+ (\alpha\beta x + \beta y +\alpha)y
\right) }{\alpha \left(
x+(\alpha x+y+1)y
\right) }},y\right).
\end{equation}
It leaves invariant the pencil of curves of genus $0$, cf. expression (\ref{inv}),
\begin{equation} \label{pencil}
 y^2 (1 + x + y) + \alpha x (\frac{x}{\beta} + x y +  y^2) = C xy,
\end{equation}
where $C$ is now a parameter distinguishing the curves in the pencil. The map $\gamma$ can be understood geometrically as the composition of two so-called $p$-switches (see \cite{KMQ} where this terminology was introduced),
\begin{equation} \label{cti}
\gamma=\iota_q\circ \iota_p.
\end{equation}
A $p$-switch $\iota_p$ maps a point $r$ on a curve of the pencil to the third point in the intersection of the curve with the line through $r$ and the involution point $p$. The pencil (\ref{pencil}) has 5 base points, in homogeneous coordinates:
\begin{equation}
(0:0:1),\
(0:-1:1),\
(1:0:0),\
(1:-1:0),\
(1:-c_1^2:0)\,,
\end{equation}
of which the first one is singular (with multiplicity 2). In formula (\ref{cti}) exactly one of $p$ or $q$ should be a non-singular base point of the pencil. The other point lies on an {\em involution curve} \cite{LA}, i.e. the involution point depends on the curve in the pencil. For example: if we take $p=(0,-1)$ (the second base point in the above list), then $q$ is
given by
\begin{equation}
q=\left({\frac { \left( C\alpha-C\beta+{\alpha}^{2}-{\beta}^{2} \right)
\alpha}{ \left( -C{\alpha}^{2}+C\alpha\,\beta-2\,{\alpha}^{3}+2\,{
\alpha}^{2}\beta+{C}^{2}+4\,C\alpha+4\,{\alpha}^{2} \right) \beta}},-{
\frac {\alpha\, \left( \alpha+\beta+C \right) }{\beta\, \left( -{
\alpha}^{2}+\alpha\,\beta+C+2\,\alpha \right) }}\right)\,,
\end{equation}
which is a parametrisation of the green curve in Figure \ref{ic}.

If we choose $p$ to be the base point at $(\infty,0)$ (the third base point in the above list), then $\iota_p$ is the so-called horizontal shift, usually denoted by $\iota_1$. We have $\gamma=\iota_r\circ \iota_1$, where
\begin{equation}
r=\left(-{\frac {{C}^{2}+C\alpha+3\,C\beta+2\,\alpha\,\beta+2\,{\beta}^{2}}{C
{\alpha}^{2}-C\alpha\,\beta+2\,{\alpha}^{2}\beta-2\,\alpha\,{\beta}^{2
}+{\alpha}^{2}-2\,\alpha\,\beta+{\beta}^{2}}},-{\frac {\alpha+\beta+C
}{C\alpha+2\,\alpha\,\beta+\alpha-\beta}}\right)\,.
\end{equation}
The map $\gamma$ is not of QRT-type (QRT-maps preserve a pencil of curves of genus 1). This is in contrast to the $3$-dimensional map obtained using periodic reduction, which is of QRT-type.

\section{Concluding remarks}\label{conclusions}
\label{sec:cr}
We introduced the notion of open boundary reductions as a new scheme to construct discrete mappings from integrable initial-boundary value problems for quad-graph systems on a strip. This represents an alternative to the well-known periodic reductions.
 One key ingredient is the idea of dual boundary equations and dual boundary consistency. This allowed us to formulate a boundary zero curvature representation, and hence prove that the discrete time evolution of the double-row monodromy matrix is an isospectral deformation (after sufficiently many iterations). The spectral functions obtained by taking the trace of the double-row monodromy matrix are invariants or $k$-invariants of the discrete $n$-dimensional mappings, where $k$ is either $2,n-1$, or $2(n-1)$.
In contrast with periodic reductions where the underlying graph forms a cylinder, open boundary reductions are built on a strip. The effect of the boundary conditions, for instance via the boundary parameter $\mu$, can be seen explicitly in the maps obtained with our method and lead to different types of maps compared to the periodic reduction method.

There are several natural continuations of the present work. A pressing question would be to investigate discrete Liouville integrability of our maps by introducing an appropriate Poisson structure. The problem of classifying integrable boundary equations for the ABS equations is still pending but the duality property seems to be a promising avenue. This is left for future work. More generally, classifying, or at least finding, integrable boundary consistency for quad equations beyond the ABS-list (e.g. Boussinesq-type quad equations) is a tantalising prospect. It would also be desirable to investigate more thoroughly the maps we obtained, as well as producing more examples, in order to make a more precise comparison with the periodic case, if at all applicable, and to establish a more precise connection with some known examples, for instance, the discrete Painlev\'e type equations.
Finally, equipped with bulk and boundary equations, initial value problems on quad-graphs with boundary can be naturally formulated, similar to how this is done for quad equations without boundary \cite{AV,VdK2}, with characteristic lines reflecting off the boundary. Some examples of well-posed initial-boundary value problems beyond the $\ZZ^2$-lattice are shown in Figures \ref{hexag} and \ref{other}. However, a general criterion of the well-posedness of initial-boundary data on generic quad-graphs with boundary still remains to be investigated.
\begin{figure}[ht]
\begin{center}
  \begin{tikzpicture}[scale=0.9]%%%%%%%%%%%%%%%%%%%%%%%%%%case 1%%%%%%%%%%%%%%%%%
    \tikzstyle{nod1}= [circle, inner sep=0pt, minimum size=3pt, draw]
    \tikzstyle{nod}= [circle, inner sep=0pt, fill=black, minimum size=3pt, draw]
    \tikzstyle{vertex}=[circle,minimum size=20pt,inner sep=0pt]
    \tikzstyle{selected vertex} = [vertex, fill=red!24]
    \tikzstyle{selected edge} = [draw,line width=1.5pt,-]
    \tikzstyle{edge} = [draw, thin,-,black] %inner lines
    \tikzstyle{ddedge} = [draw, densely dotted,-,black] %dotted lines
    \tikzstyle{dedge} = [draw, dashed,-,black] %dashed lines
    \tikzstyle{eedge} = [draw, thick,-,black] %outer lines

    \pgfmathsetmacro \d {1.1}% inner lenght
    \pgfmathsetmacro \cd {0.5}% inner lenght
    \pgfmathsetmacro \dx {{\d*sin(30)}}
    \pgfmathsetmacro \dy {{\d*sin(60)}}
    \pgfmathsetmacro \cdx {{\cd*sin(30)}}
    \pgfmathsetmacro \cdy {{\cd*sin(60)}}

    % 1st floor v, v1, v4, v14
    \node[nod1] (x0) at (0,0) {};
    \node[nod] (x1) at (\dx,\dy) {};
    \node[nod] (x2) at (0,2*\dy) {};
    \node[nod1] (x3) at (\dx,3*\dy) {};
    \node[nod1] (x4) at (0,4*\dy) {};
    \node[nod1] (x5) at (\dx,5*\dy) {};
    \node[nod1] (x6) at (0,6*\dy) {};
    \coordinate (bx0) at (\cdx,-\cdy);
    \coordinate (tx6) at (\cdx,6*\dy+\cdy);

    \draw (x0) -- (x1) -- (x2) -- (x3) node[left] {\footnotesize$x_2'$}-- (x4) node[left] {\footnotesize$x_1'$} -- (x5) --  (x6) ;
    \draw (x0) -- (x4);
    \draw (x4) -- (x6);
    \draw[dashed] (x0) -- (bx0);
    \draw[dashed] (x6) -- (tx6);

    \node[nod1] (z0) at (2*\dx+\d,0) {};
    \node[nod] (z1) at (\dx+\d,\dy) {};
    \node[nod] (z2) at (2*\dx+\d,2*\dy) {};
    \node[nod1] (z3) at (\dx+\d,3*\dy) {};
    \node[nod1] (z4) at (2*\dx+\d,4*\dy) {};
    \node[nod1] (z5) at (\dx+\d,5*\dy) {};
    \node[nod1] (z6) at (2*\dx+\d,6*\dy) {};
    \coordinate (bz0) at (2*\dx+\d-\cdx,-\cdy);
    \coordinate (tz6) at (2*\dx+\d-\cdx,6*\dy+\cdy);

    \draw (z0) -- (z1) -- (z2) -- (z3)  node[right] {\footnotesize$x_3'$}-- (z4) node[right] {\footnotesize$x_4'$} -- (z5) --  (z6) ;
    \draw (z0) -- (z4);
    \draw (z4) -- (z6);
    \draw[dashed] (z0) -- (bz0);
    \draw[dashed] (z6) -- (tz6);

    \node[nod1] (y0) at (\d,0) {};
    \node[nod1] (y2) at (\d,2*\dy) {};
    \node[nod1] (y4) at (\d,4*\dy) {};
    \node[nod1] (y6) at (\d,6*\dy) {};
    \coordinate (by0) at (\d+\cdx,-\cdy);
    \coordinate (ty6) at (\d+\cdx,6*\dy+\cdy);
    \draw[dashed] (y0) -- (by0);
    \draw[dashed] (y6) -- (ty6);

    \draw (y0) -- (z1) -- (y2) -- (z3) -- (y4) -- (z5) --  (y6) ;	
    \draw (x0) -- (y0);
    \draw (x2) -- (y2);
    \draw (x4) -- (y4);
    \draw (x6) -- (y6);
    \draw (z1) -- (x1);
    \draw (z3) -- (x3);
    \draw (z5) -- (x5);
  \draw[thick] (x2) node[left] {\footnotesize$x_1$} -- (x1)node[left] {\footnotesize$x_2$} -- (z1)node[right] {\footnotesize$x_3$} -- (z2)node[right] {\footnotesize$x_4$};
  \end{tikzpicture}  \hspace{1.5cm}
\begin{tikzpicture}[scale=0.9] %%%%%%%%%%%%%%%%%%%%% case 2

    \tikzstyle{nod1}= [circle, inner sep=0pt, minimum size=3pt, draw]
    \tikzstyle{nod}= [circle, inner sep=0pt, fill=black, minimum size=3pt, draw]
    \tikzstyle{vertex}=[circle,minimum size=20pt,inner sep=0pt]
    \tikzstyle{selected vertex} = [vertex, fill=red!24]
    \tikzstyle{selected edge} = [draw,line width=1.5pt,-]
    \tikzstyle{edge} = [draw, thin,-,black] %inner lines
    \tikzstyle{ddedge} = [draw, densely dotted,-,black] %dotted lines
    \tikzstyle{dedge} = [draw, dashed,-,black] %dashed lines
    \tikzstyle{eedge} = [draw, thick,-,black] %outer lines

    \pgfmathsetmacro \d {1.1}% inner lenght
    \pgfmathsetmacro \cd {0.5}% inner lenght
    \pgfmathsetmacro \dx {{\d*sin(30)}}
    \pgfmathsetmacro \dy {{\d*sin(60)}}
    \pgfmathsetmacro \cdx {{\cd*sin(30)}}
    \pgfmathsetmacro \cdy {{\cd*sin(60)}}

    % 1st floor v, v1, v4, v14
    \node[nod1] (x0) at (0,0) {};
    \node[nod] (x1) at (\dx,\dy) {};
    \node[nod] (x2) at (0,2*\dy) {};
    \node[nod1] (x3) at (\dx,3*\dy) {};
    \node[nod1] (x4) at (0,4*\dy) {};
    \node[nod1] (x5) at (\dx,5*\dy) {};
    \node[nod1] (x6) at (0,6*\dy) {};
    \coordinate (bx0) at (\cdx,-\cdy);
    \coordinate (tx6) at (\cdx,6*\dy+\cdy);

    \draw (x0) -- (x1) -- (x2) -- (x3) node[left] {\footnotesize$x_2'$}-- (x4) node[left] {\footnotesize$x_1'$}-- (x5) --  (x6) ;
    \draw (x0) -- (x4);
    \draw (x4) -- (x6);
    \draw[dashed] (x0) -- (bx0);
    \draw[dashed] (x6) -- (tx6);

    \node[nod1] (z0) at (2*\dx+\d,0) {};
    \node[nod] (z1) at (\dx+\d,\dy) {};
    \node[nod] (z2) at (2*\dx+\d,2*\dy) {};
    \node[nod1] (z3) at (\dx+\d,3*\dy) {};
    \node[nod1] (z4) at (2*\dx+\d,4*\dy) {};
    \node[nod1] (z5) at (\dx+\d,5*\dy) {};
    \node[nod1] (z6) at (2*\dx+\d,6*\dy) {};
    \coordinate (bz0) at (2*\dx+\d-\cdx,-\cdy);
    \coordinate (tz6) at (2*\dx+\d-\cdx,6*\dy+\cdy);

    \draw (z0) -- (z1) -- (z2) -- (z3)  node[right] {\footnotesize$x_3'$}-- (z4) node[right] {\footnotesize$x_4'$} -- (z5) --  (z6) ;
    \draw (z0) -- (z4);
    \draw (z4) -- (z6);
    \draw[dashed] (z0) -- (bz0);
    \draw[dashed] (z6) -- (tz6);

    \node[nod1] (y0) at (\d,0) {};
    \node[nod1] (y2) at (\d,2*\dy) {};
    \node[nod1] (y4) at (\d,4*\dy) {};
    \node[nod1] (y6) at (\d,6*\dy) {};

        \coordinate (by0) at (\d-\cdx,-\cdy);
    \coordinate (ty6) at (\d-\cdx,6*\dy+\cdy);
    \draw[dashed] (y0) -- (by0);
    \draw[dashed] (y6) -- (ty6);

    \draw (y0) -- (x1) -- (y2) -- (x3) -- (y4) -- (x5) --  (y6) ;	
    \draw (z0) -- (y0);
    \draw (z2) -- (y2);
    \draw (z4) -- (y4);
    \draw (z6) -- (y6);
    \draw (x1) -- (z1);
    \draw (x3) -- (z3);
    \draw (x5) -- (z5);

    \draw[thick] (x2) node[left] {\footnotesize$x_1$} -- (x1)node[left] {\footnotesize$x_2$} -- (z1)node[right] {\footnotesize$x_3$} -- (z2)node[right] {\footnotesize$x_4$};

  \end{tikzpicture} \hspace{1.5cm}
    \begin{tikzpicture}[scale=0.9]%%%%%%%%%%%%%%%% case 3

    \tikzstyle{nod1}= [circle, inner sep=0pt, minimum size=3pt, draw]
    \tikzstyle{nod}= [circle, inner sep=0pt, fill=black, minimum size=3pt, draw]
    \tikzstyle{vertex}=[circle,minimum size=20pt,inner sep=0pt]
    \tikzstyle{selected vertex} = [vertex, fill=red!24]
    \tikzstyle{selected edge} = [draw,line width=1.5pt,-]
    \tikzstyle{edge} = [draw, thin,-,black] %inner lines
    \tikzstyle{ddedge} = [draw, densely dotted,-,black] %dotted lines
    \tikzstyle{dedge} = [draw, dashed,-,black] %dashed lines
    \tikzstyle{eedge} = [draw, thick,-,black] %outer lines

    \pgfmathsetmacro \d {1.1}% inner lenght
    \pgfmathsetmacro \cd {0.5}% inner lenght
    \pgfmathsetmacro \dx {{\d*sin(30)}}
    \pgfmathsetmacro \dy {{\d*sin(60)}}
    \pgfmathsetmacro \cdx {{\cd*sin(30)}}
    \pgfmathsetmacro \cdy {{\cd*sin(60)}}

    % 1st floor v, v1, v4, v14
    \node[nod1] (x0) at (0,0) {};
    \node[nod] (x1) at (\dx,\dy) {};
    \node[nod] (x2) at (0,2*\dy) {};
    \node[nod1] (x3) at (\dx,3*\dy) {};
    \node[nod1] (x4) at (0,4*\dy) {};
    \node[nod1] (x5) at (\dx,5*\dy) {};
    \node[nod1] (x6) at (0,6*\dy) {};
    \coordinate (bx0) at (\cdx,-\cdy);
    \coordinate (tx6) at (\cdx,6*\dy+\cdy);

    \draw (x0) -- (x1) -- (x2) -- (x3) node[left] {\footnotesize$x_2'$}-- (x4) node[left] {\footnotesize$x_1'$}-- (x5) --  (x6) ;
    \draw (x0) -- (x4);
    \draw (x4) -- (x6);
    \draw[dashed] (x0) -- (bx0);
    \draw[dashed] (x6) -- (tx6);

    \node[nod1] (z0) at (2*\dx+\d,0) {};
    \node[nod] (z1) at (\dx+\d,\dy) {};
    \node[nod] (z2) at (2*\dx+\d,2*\dy) {};
    \node[nod1] (z3) at (\dx+\d,3*\dy) {};
    \node[nod1] (z4) at (2*\dx+\d,4*\dy) {};
    \node[nod1] (z5) at (\dx+\d,5*\dy) {};
    \node[nod1] (z6) at (2*\dx+\d,6*\dy) {};
    \coordinate (bz0) at (2*\dx+\d-\cdx,-\cdy);
    \coordinate (tz6) at (2*\dx+\d-\cdx,6*\dy+\cdy);

    \draw (z0) -- (z1) -- (z2) -- (z3)  node[right] {\footnotesize$x_3'$}-- (z4) node[right] {\footnotesize$x_4'$} -- (z5) --  (z6) ;
    \draw (z0) -- (z4);
    \draw (z4) -- (z6);
    \draw[dashed] (z0) -- (bz0);
    \draw[dashed] (z6) -- (tz6);

    \node[nod1] (y0) at (\d,0) {};
    \node[nod1] (y2) at (\d,2*\dy) {};
    \node[nod1] (y4) at (\d,4*\dy) {};
    \node[nod1] (y6) at (\d,6*\dy) {};
        \coordinate (by0) at (\d-\cdx,-\cdy);
    \coordinate (ty6) at (\d+\cdx,6*\dy+\cdy);
    \draw[dashed] (y0) -- (by0);
    \draw[dashed] (y6) -- (ty6);

    \draw (y0) -- (x1);
    \draw (y6) -- (z5);
    \draw (z1) -- (y2) -- (z3);
    \draw (x3) -- (y4) -- (x5);
    \draw (z0) -- (y0);
    \draw (x1) -- (z1);
    \draw (x2) -- (y2);
    \draw (x3) -- (z3);
    \draw (y4) -- (z4);
    \draw (x5) -- (z5);
    \draw (x6) -- (y6);

    \draw[thick] (x2) node[left] {\footnotesize$x_1$} -- (x1)node[left] {\footnotesize$x_2$} -- (z1)node[right] {\footnotesize$x_3$} -- (z2)node[right] {\footnotesize$x_4$};
      \end{tikzpicture}
\end{center}
\caption{\label{hexag} Open boundary reduction on a hexagonal lattice. The three configurations amount to the same map due to the $3$D-consistency of the bulk equation. }\label{GL1}
\end{figure}
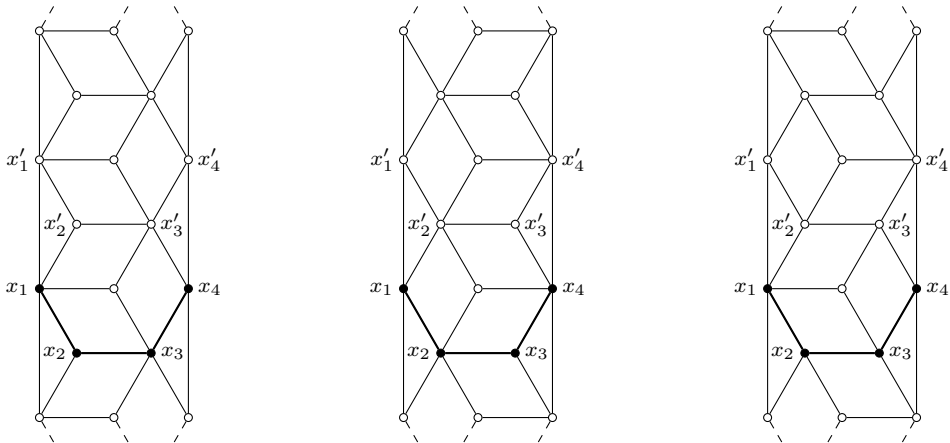

\begin{figure}[ht]
\begin{center}
  \begin{tikzpicture}[scale=0.9]%%%%%%%%%%%%%%%%%%%%%%%%%%case 4%%%%%%%%%%%%%%%%%
    \tikzstyle{nod1}= [circle, inner sep=0pt, minimum size=3pt, draw]
    \tikzstyle{nod}= [circle, inner sep=0pt, fill=black, minimum size=3pt, draw]
    \tikzstyle{vertex}=[circle,minimum size=20pt,inner sep=0pt]
    \tikzstyle{selected vertex} = [vertex, fill=red!24]
    \tikzstyle{selected edge} = [draw,line width=1.5pt,-]
    \tikzstyle{edge} = [draw, thin,-,black] %inner lines
    \tikzstyle{ddedge} = [draw, densely dotted,-,black] %dotted lines
    \tikzstyle{dedge} = [draw, dashed,-,black] %dashed lines
    \tikzstyle{eedge} = [draw, thick,-,black] %outer lines

    \pgfmathsetmacro \d {1.1}% inner lenght
    \pgfmathsetmacro \cd {0.5}% inner lenght
    \pgfmathsetmacro \dx {{\d*sin(30)}}
    \pgfmathsetmacro \dy {{\d*sin(60)}}
    \pgfmathsetmacro \cdx {{\cd*sin(30)}}
    \pgfmathsetmacro \cdy {{\cd*sin(80)}}

    % 1st floor v, v1, v4, v14
    \node[nod1] (x0) at (0,0) {};
    \node[nod] (x1) at (\dx,\dy) {};
    \node[nod] (x2) at (0,2*\dy) {};
    \node[nod1] (x3) at (\dx,3*\dy) {};
    \node[nod1] (x4) at (0,4*\dy) {};
    \coordinate (bx0) at (\cdx,-\cdy);
    \coordinate (tx4) at (\cdx,4*\dy+\cdy);
    \node[nod1] (z0) at (0+2*\d,0) {};
    \node[nod] (z1) at (\dx+2*\d,\dy) {};
    \node[nod1] (z2) at (0+2*\d,2*\dy) {};
    \node[nod1] (z3) at (\dx+2*\d,3*\dy) {};
    \node[nod1] (z4) at (0+2*\d,4*\dy) {};
    \coordinate (bz0) at (\cdx+2*\d,-\cdy);
    \coordinate (tz4) at (\cdx+2*\d,4*\dy+\cdy);

    \node[nod1] (y0) at (0+\d,0) {};
    \node[nod] (y1) at (\dx+\d,\dy) {};
    \node[nod1] (y2) at (0+\d,2*\dy) {};
    \node[nod1] (y3) at (\dx+\d,3*\dy) {};
    \node[nod1] (y4) at (0+\d,4*\dy) {};
    \coordinate (by0) at (\cdx+\d,-\cdy);
    \coordinate (ty4) at (\cdx+\d,4*\dy+\cdy);

    \draw (x0) -- (x1) node[left] {\footnotesize$x_2$}-- (x2)node[left] {\footnotesize$x_1$} -- (x3) node[left] {\footnotesize$x_2'$}-- (x4) node[left] {\footnotesize$x_1'$}  ;
    \draw (z0) -- (z1)  node[right] {\footnotesize$x_4$}-- (z2) -- (z3) node[right] {\footnotesize$x_4'$}-- (z4)   ;
    \draw (x0) -- (x4);
    \draw (\dx+2*\d, 0 ) -- (\dx+2*\d, 4*\dy);
    \draw[dashed] (\dx+2*\d, 0 ) -- (\dx+2*\d, -\cdy);
    \draw[dashed] (\dx+2*\d, 4*\dy ) -- (\dx+2*\d, 4*\dy+\cdy);
     \draw[dashed] (x0) -- (bx0);
     \draw[dashed] (x4) -- (tx4);
     \draw[dashed] (z0) -- (bz0);
     \draw[dashed] (z4) -- (tz4);
     \draw[dashed] (y0) -- (by0);
     \draw[dashed] (y4) -- (ty4);
\draw[thick] (x2) -- (x1) -- (y1)node[below right] {\footnotesize$x_3$} -- (z1) ;
\draw  (x3) -- (y3)node[below right] {\footnotesize$x_3'$} -- (z3) ;
\draw  (x2) -- (y2) -- (z2) ;
\draw  (x0) -- (y0) -- (z0) ;
\draw  (x4) -- (y4) -- (z4) ;
\draw  (y0) -- (y1) -- (y2)-- (y3) -- (y4) ;
\end{tikzpicture}  \hspace{1.5cm}
\begin{tikzpicture}[scale=0.9] %%%%%%%%%%%%%%%%%%%%% case 2

    \tikzstyle{nod1}= [circle, inner sep=0pt, minimum size=3pt, draw]
    \tikzstyle{nod}= [circle, inner sep=0pt, fill=black, minimum size=3pt, draw]
    \tikzstyle{vertex}=[circle,minimum size=20pt,inner sep=0pt]
    \tikzstyle{selected vertex} = [vertex, fill=red!24]
    \tikzstyle{selected edge} = [draw,line width=1.5pt,-]
    \tikzstyle{edge} = [draw, thin,-,black] %inner lines
    \tikzstyle{ddedge} = [draw, densely dotted,-,black] %dotted lines
    \tikzstyle{dedge} = [draw, dashed,-,black] %dashed lines
    \tikzstyle{eedge} = [draw, thick,-,black] %outer lines

    \pgfmathsetmacro \d {1.1}% inner lenght
    \pgfmathsetmacro \cd {0.5}% inner lenght
    \pgfmathsetmacro \dx {{\d*sin(30)}}
    \pgfmathsetmacro \dy {{\d*sin(60)}}
    \pgfmathsetmacro \cdx {{\cd*sin(30)}}
    \pgfmathsetmacro \cdy {{\cd*sin(70)}}

    % 1st floor v, v1, v4, v14
    \node[nod1] (x0) at (0,0) {};
    \node[nod] (x1) at (\dx,\dy) {};
    \node[nod] (x2) at (0,2*\dy) {};
    \node[nod1] (x3) at (\dx,3*\dy) {};
    \node[nod1] (x4) at (0,4*\dy) {};
    \coordinate (bx0) at (\cdx,-\cdy);
    \coordinate (tx4) at (\cdx,4*\dy+\cdy);

    \node[nod1] (y0) at (\dy,\dx) {};
    \node[nod1] (z0) at (\d+\dy,0) {};
    \node[nod] (y1) at (\dy,\dy) {};
    \node[nod] (z1) at (2*\dy+\dx,\dy) {};
    \node[nod1] (y2) at (\dy,2*\dy-\dx) {};
    \node[nod1] (z2) at (\d+\dy,2*\dy) {};
    \node[nod1] (z3) at (2*\dy+\dx,3*\dy) {};
    \node[nod1] (z4) at (\d+\dy,4*\dy) {};
    \node[nod1] (y3) at (\dy,2*\dy+\dx) {};
    \node[nod1] (y4) at (0+\dy,3*\dy) {};
    \node[nod1] (y5) at (0+\dy,4*\dy-\dx) {};

    % \coordinate (by0) at (\cdx+\d,-\cdy);
    % \coordinate (ty4) at (\cdx+\d,4*\dy+\cdy);

 \draw (x0) -- (x1) node[left] {\footnotesize$x_2$}-- (x2)node[left] {\footnotesize$x_1$} -- (x3) node[left] {\footnotesize$x_2'$}-- (x4) node[left] {\footnotesize$x_1'$}  ;
     \draw (z0) -- (z1)  node[right] {\footnotesize$x_4$}-- (z2) -- (z3) node[right] {\footnotesize$x_4'$}-- (z4)   ;
     \draw (x0) -- (x4);
     \draw[dashed] (x0) -- (bx0);
     \draw[dashed]   (x0) -- (\cdy,-\cdx);
     \draw[dashed] (x4) -- (tx4);
     \draw[dashed]   (x4) -- (\cdy,4*\dy+\cdx);
     \draw[dashed]   (z4) --  (\d+\dy-\cdy,4*\dy+\cdx);
     \draw[dashed]   (z4) --  (\d+\dy+\cdx,4*\dy+\cdy);
     \draw[dashed]   (z0) --  (\d+\dy-\cdy,-\cdx);
     \draw[dashed]   (z0) --  (\d+\dy+\cdx,0-\cdy);
     \draw ( 2*\dy+\dx,0) --  (z3)-- ( 2*\dy+\dx,4*\dy);
    \draw (x0) -- (y0)-- (z0);
      \draw (x1) -- (z1);
      \draw (y0) -- (y2) ;
      \draw (x2) -- (y2) -- (z2) -- (y3) -- (x2);
      \draw (x4) -- (y5) -- (z4);
      \draw (x3) -- (y4)-- (z3);
      \draw (y3) -- (y4) node[above right] {\footnotesize$x_3'$}-- (y5);
 \draw[thick] (x2) -- (x1) -- (y1)node[above right] {\footnotesize$x_3$} -- (z1) ;
    \draw[dashed] (\dx+2*\dy, 0 ) -- (\dx+2*\dy, -\cdy);
    \draw[dashed] (\dx+2*\dy, 4*\dy ) -- (\dx+2*\dy, 4*\dy+\cdy);

  \end{tikzpicture}\end{center}
\caption{\label{other} Other possibilities of open boundary reductions beyond $\ZZ^2$-lattice.}\label{GL2}
\end{figure}
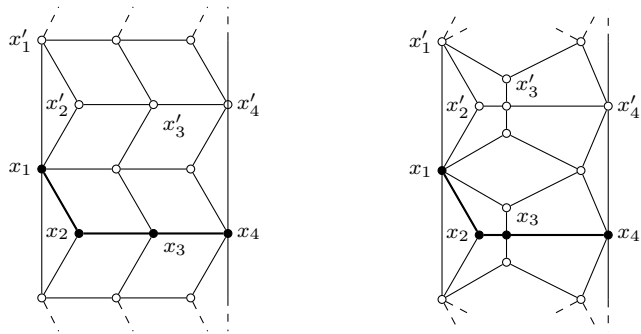

\subsection*{Acknowledgements}
This project is supported by NSFC (No.~11875040), CSC (No.~201906895010), and La Trobe University China Strategy Implementation and China Seed Funding grants. % This work was initiated when all three authors attended the Asymptotic, Algebraic and Geometric Aspects of Integrable Systems Workshop at the Tsinghua Sanya Intermational Mathemtaical Forum, 9-13 April 2018.

\end{document}